\documentclass[twocolumn,showpacs,prl,aps,10pt,superscriptaddress]{revtex4-1}
\usepackage{amssymb}
\usepackage{amsmath}
\usepackage{graphicx}
\usepackage{color,comment}

\vbadness=\maxdimen

\begin{document}

\title{Bipartite and tripartite output entanglement in 3-mode optomechanical
systems}
\author{Ying-Dan Wang}
\affiliation{CEMS, RIKEN, Wako, Saitama 351-0198, Japan}
\affiliation{Institute of Theoretical Physics, Chinese Academy of Sciences, Beijing 100190, China}
\affiliation{Department of Physics, McGill University, 3600 rue University, Montreal, QC
Canada H3A 2T8}
\author{Stefano Chesi}
\affiliation{CEMS, RIKEN, Wako, Saitama 351-0198, Japan}
\affiliation{Beijing Computational Science Research Center, Beijing 100084, China}
\author{Aashish~A.~Clerk}
\affiliation{Department of Physics, McGill University, 3600 rue University, Montreal, QC
Canada H3A 2T8}
\date{\today}

\begin{abstract}
We provide analytic insight into the generation of stationary itinerant photon
entanglement in a 3-mode optomechanical system. We identify the parameter
regime of maximal entanglement, and show that strong entanglement is
possible even for weak many-photon optomechanical couplings. We also show that strong
tripartite entanglement is generated between the photonic and phononic output fields; 
unlike the bipartite photon-photon entanglement,
this tripartite entanglement diverges as one approaches the boundary of
system stability. 
\end{abstract}


\maketitle



Entanglement is one of the most fascinating and potentially useful
aspects of quantum systems. Of particular interest is the generation of
entangled itinerant quanta (which can be easily spatially separated), and of
true multipartite entanglement (involving irreducible correlations between
three or more subsystems). These goals have been the subject of considerable
theoretical and experimental work, in a variety of systems spanning quantum
optics setups~\cite{Aoki2003,Pan2012}, cold atoms~\cite{Raimond2001}, superconducting
circuits~\cite{Neeley2010,Dicarlo2010,Flurin2012} and spin qubits~\cite%
{Bernien2013}.  Optomechanical systems \cite{FlorianRMP}, where mechanical motion
interacts with electromagnetic fields, could be another powerful platform to
realize these goals. A key advantage here is the potential to use mechanical
motion to entangle disparate subsystems (e.g.~microwave and optical
photons). A number of schemes to generate entangled photons in optomechanics
have been studied theoretically ~\cite%
{Paternostro2007,Genes2008,Wipf2008,Hofer2011,Barzanjeh2012,Tian2013,Kuzyk2013}. Recent experiments
have also demonstrated mechanically-mediated entanglement between two
microwave pulses~\cite{Palomaki2013}. 

\begin{figure}[t]
\begin{center}
\raisebox{0pt}{\includegraphics[width=0.4\textwidth]{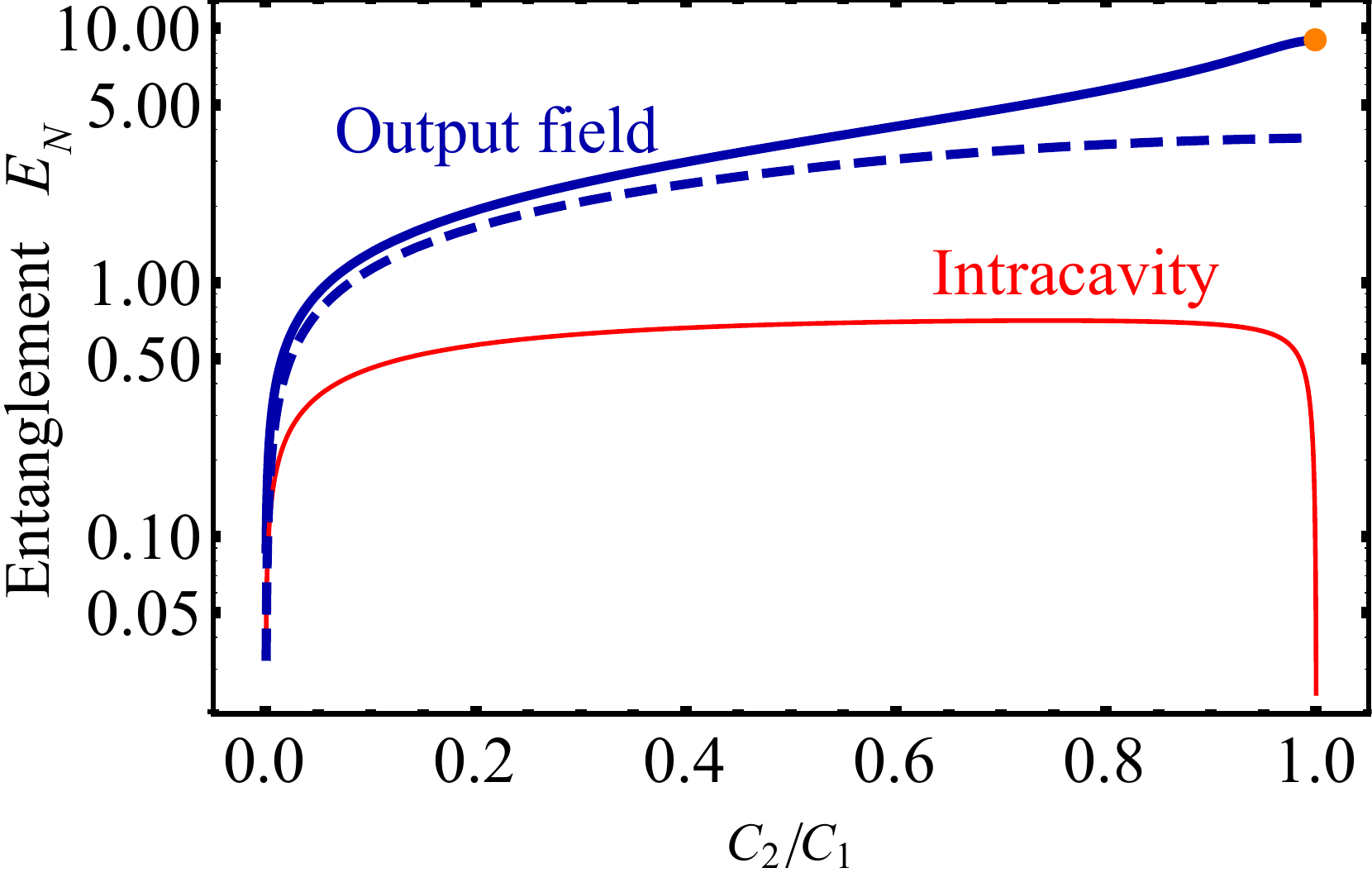}} %
\raisebox{1.4cm}[0cm][0cm]{\hspace{1.2cm}%
\includegraphics[width=5cm]{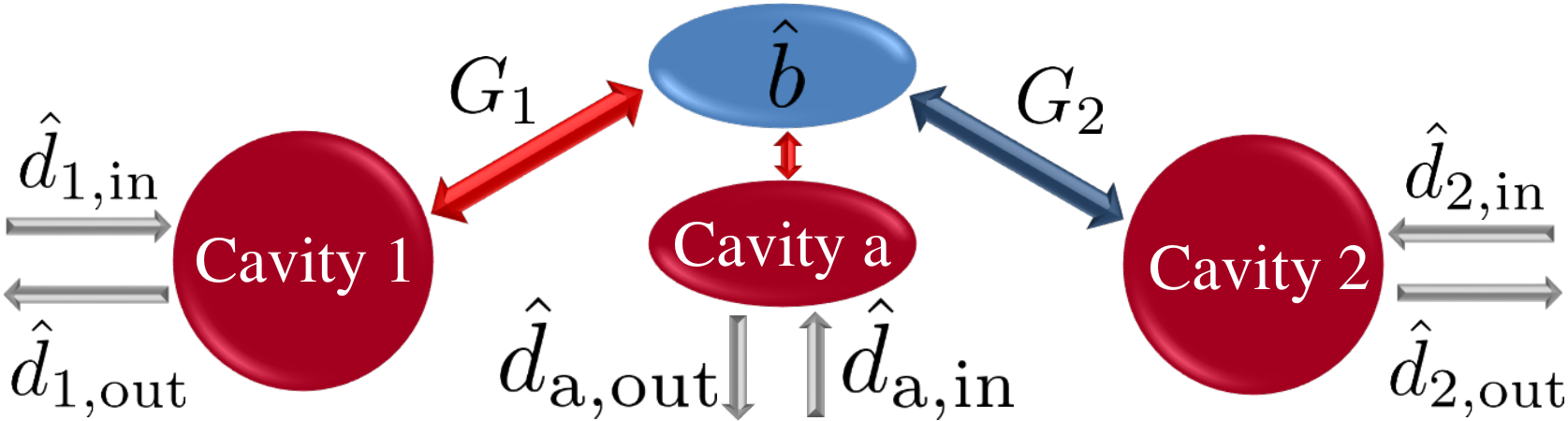}} \vspace{-0.7cm} 
\end{center}
\caption{Inset: Two driven cavities interact with a common mechanical
resonator, generating entanglement in the optical outputs. An auxiliary third
cavity can be used to cavity cool the mechanics, and to make the mechanical
output mode accessible. Main: entanglement of the two cavity output fields
on resonance (i.e.~$\omega =0$) (thick blue) and corresponding
intracavity entanglement (thin red) as functions of $C_2$, with $C_1 =
4000 $ and $\omega_\mathrm{m} \gg \kappa$ (allowing use of the RWA).
The orange dot denotes the value of $\ln 2C_1$ as in Eq.~(\ref{maxen}%
) with $\kappa_1 \geq \protect\kappa_2$. Results are for zero
temperature, except for the blue dashed line (mechanical bath occupancy $%
N_\mathrm{m}=100$).  The output $E_N$ curves only depend on the $C_i$.  For the intracavity curve (thin red), we have also assumed strong coupling, taking $G_1:\kappa:\protect\gamma=100:10:1$.}
\label{fig:inputvsoutput}
\end{figure}

Here, we analyze theoretically both itinerant and multipartite entanglement
in a 3-mode optomechanical system where two cavities are coupled to a single
mode of a mechanical resonator (see inset of Fig.~\ref{fig:inputvsoutput}).
This setup  has been realized in several recent experiments~\cite{Dong2012,Hill2012,Andrews2014}. 
Previous theory work examined bipartite output entanglement in this system largely numerically~\cite{Wipf2008, Barzanjeh2012,Tian2013,Kuzyk2013} ,
focusing on experimentally-challenging strong-coupling regimes~\cite{Barzanjeh2012,Tian2013} or on transient regimes \cite{Kuzyk2013}.
In contrast, we focus here on generating stationary output entanglement with weak many-photon optomechanical couplings.
We provide a complete yet simple \textit{analytic} understanding of the
physics. This allows us to illustrate the trade-off between large
entanglement and thermal resilience, as well as to identify the parameter
regime of maximum entanglement, a regime which corresponds to a simple
matching of optomechanical cooperativities. Surprisingly, this condition
coincides with the least favorable regime for the generation of intra-cavity
entanglement.  We also show that entanglement is optimal between time-delayed pairs of
wavepackets. 

We also address the generation of tripartite entanglement in such a hybrid
3-mode system, considering correlations between both output photons and
phonons. While usually ignored, the mechanical output field could be
accessed experimentally, using for example optomechanical crystal geometries
with phonon waveguides~\cite{Safavi2011}, or by having the mechanical
dissipation be dominated by a third auxiliary cavity used for cooling~\cite{EPAPS}.
We find that true hybrid tripartite
entanglement is indeed created: the output state corresponds to a
\textquotedblleft twice-squeezed vacuum", involving the action of two
2-mode squeezing operations involving all three modes.  We also quantify
this entanglement using the Gaussian R\'enyi-2 measure~\cite{Adesso2012}.
Besides being of fundamental interest, such tripartite entangled states have applications to
a variety of quantum information processing tasks such as teleportation,
tele-cloning and dense coding~\cite{Braunstein2005,Furusawa2006}. While our emphasis here
is on optomechanics, our results also apply to other bosonic 3-mode systems
(as could be realized, e.g., with superconducting circuits~\cite%
{Bergeal2010,Baust2014}). The present setup is especially suited to the
continous generation of non-local multipartite entanglement, as the phonons
and photons from the two cavities are all emitted into spatially separated
outputs. Note that Genes et al.~\cite{Genes2008} also studied tri-partite entanglement in
an optomechanical system, though in a setting where the entangled subsystems were not all spatially separated or itinerant.

\textit{System and instabilities --} The Hamiltonian $\hat{H}=\omega _{\mathrm{%
m}}\hat{b}^{\dag }\hat{b}+\sum_{i=1,2}\left[ \omega _{i}\hat{a}_{i}^{\dag }%
\hat{a}_{i}+g_{i}\left( \hat{b}^{\dag }+\hat{b}\right) \hat{a}_{i}^{\dag }%
\hat{a}_{i}\right] +\hat{H}_{\mathrm{diss}}$ governs the system's dynamics,
with $\hat{a}_{i}$ the annihilation operator for cavity $i$ (frequency $%
\omega _{i}$, damping rate $\kappa _{i}$), $\hat{b}$ the annihilation
operator of the mechanical mode (frequency $\omega _{\mathrm{m}}$, damping
rate $\gamma $), and $g_{i}$ the optomechanical coupling strengths. $\hat{H}%
_{\mathrm{diss}}$ describes the dissipation of each mode, treated via
standard input-output theory \cite{Gardinerbook}. In order to generate
steady-state entanglement, cavity $1$ ($2$) is driven at the red (blue)
sideband associated with the mechanical resonator: $\omega _{d1}=\omega
_{1}-\omega _{\mathrm{m}}$ and $\omega _{d2}=\omega _{2}+\omega _{\mathrm{m}%
} $. By working in an interaction picture with respect to the free
Hamiltonian and following the usual linearization procedure 
\cite{FlorianRMP}, we have $\hat{H}^{\mathrm{R}}=\hat{H}_{\mathrm{int}}+\hat{H}%
_{\mathrm{CR}}\left( t\right) +\hat{H}_{\mathrm{diss}}$ with 
\begin{equation}
\hat{H}_{\mathrm{int}}=\left( G_{1}\hat{b}^{\dag }\hat{d}_{1}+G_{2}\hat{b}%
\hat{d}_{2}\right) +h.c.=\tilde{G}\left( \hat{b}^{\dag }\hat{\beta}%
_{A}+h.c.\right) ,  \label{eq:Hint}
\end{equation}%
and $\hat{H}_{\mathrm{CR}}\left( t\right) =\tilde{G}\left( \hat{b}^{\dag }%
\hat{\beta}_{A}^{\dag }e^{2i\omega _\mathrm{m}t}+h.c.\right) $. Here $\hat{d}_{i}=%
\hat{a}_{i}-\bar{a}_{i}$ with $\bar{a}_{i}$ the classical cavity amplitude. $%
G_{i}=g_{i}\bar{a}_{i}$ is the dressed coupling (we take $g_{i},\bar{a}%
_{i}>0 $ without loss of generality), $\tilde{G}=\sqrt{G_{1}^{2}-G_{2}^{2}}$
and $\hat{\beta}_{A}=\hat{d}_{1}\cosh r+\hat{d}_{2}^{\dag }\sinh r$ ($%
r=\tanh ^{-1}\left( G_{2}/G_{1}\right) $) is a Bogoliubov mode.

We first focus on the resolved-sideband regime $\omega _{\mathrm{m}}\gg
\kappa _{1},\kappa _{2} $ and make the rotating wave approximation (RWA) by
neglecting $\hat{H}_{\mathrm{CR}}$ (see \cite{EPAPS} for non-RWA
corrections).The combined swapping and entangling interactions in $\hat{H}_{\mathrm{int}}$
lead to a net entangling interaction between the two cavity modes; as
discussed in~\cite{Wang2013}, this interaction has a
fundamentally dissipative nature. It is useful to diagonalize $\hat{H}_{%
\mathrm{int}}$ in terms of three normal modes \cite{Wang2013, Tian2013}: one
``mechanically-dark" Bogoliubov mode $\hat{\beta}_{B}=\hat{d}_{1}\sinh r+%
\hat{d}_{2}^{\dag }\cosh r$ (which is robust to mechanical thermal noise),
and two coupled eigenmodes involving both the mechanics and cavities.

Given the blue-detuned laser drive, a first question involves the stability
of our linearized system. The Routh-Hurwitz conditions~\cite{DeJesus1987} yields two requirements~\cite{Tian2013,Wang2013} to guarantee stability. The
first is that $\gamma _{\mathrm{tot}}>0$, where $\gamma _{ \mathrm{tot}%
}=\gamma +4G_{1}^{2}/\kappa _{1}-4G_{2}^{2}/\kappa _{2}$ is the effective
damping rate of the mechanical resonator interacting with the two cavities.
Focusing on the interesting and relevant regime of strong cooperativities $%
C_{i} \equiv 4G_{i}^{2}/ \left( \gamma \kappa _{i}\right) \gg 1 $ and $%
\kappa_i \gg \gamma$, the two requirements can be combined into: 
\begin{equation}
G_{1}^{2}/G_{2}^{2}>\max (\kappa _{2}/\kappa _{1},\kappa _{1}/\kappa _{2}).
\label{stability}
\end{equation}


\textit{Cavity output entanglement -- } We start by considering the
entanglement of light leaving the two cavities. The frequency-resolved
output modes $\hat{d}_{i}^{\mathrm{out}}\left[ \omega \right] \equiv \int
d\omega e^{i\omega t}\hat{d}_{i}^{\mathrm{out}}\left( t\right) /\sqrt{2\pi }$
are related to the input by $\hat{d}_{i}^{\mathrm{out}}\left[ \omega \right]
=\sum_{j=1}^3S_{ij}[ \omega ] \hat{d}_{j}^{\mathrm{in}}\left[
\omega \right] $ ($j=3$ denotes the mechanical fields), where the scattering
matrix $\mathbf{S}[\omega]$ is obtained
straightforwardly from the system Langevin
equations (with RWA) and input-output relations, see Eq.~(S1) in~\cite{EPAPS}.

For simplicity, we consider output temporal modes in a
bandwidth $\sigma $ centered about the frequency $\omega$ described by the following
canonical mode operators 
\begin{equation}
\hat{D}_{i}^{\mathrm{out}}\left[ \omega ,\sigma ,\tau _{i}\right] =\frac{1}{%
\sqrt{\sigma }}\int_{\omega -\frac{\sigma }{2}}^{\omega +\frac{\sigma }{2}%
}d\omega ^{\prime }e^{-i\omega ^{\prime }\tau _{i}}\hat{d}_{i}^{\mathrm{out%
}}\left[ \omega ^{\prime }\right] .
\label{eq:DMode}
\end{equation}%
Here, $\tau _{i}$ sets the absolute time at which the wavepacket of interest
is emitted from cavity $i$; without loss of generality, we set $\tau _{2}=0$. 
The two-mode entanglement can be quantified using the logarithmic
negativity  
$E_{N}^{\mathrm{out}}\left[ \omega ,\sigma ,\tau _{1}\right] $ \cite%
{Vidal2002,Plenio2005}. For clarity, we start by discussing the case $\sigma
\rightarrow 0$; the result is then independent of $\tau _{1}$ ($\hat{D}_{i}^{\mathrm{out}}%
\left[ \omega \right] \equiv \hat{D}_{i}^{\mathrm{out}}\left[ \omega ,0,\tau _{i}\right] e^{i\omega\tau_i} $, 
$E_{N}^{\mathrm{out}}\left[\omega \right] \equiv E_{N}^{\mathrm{out}}\left[ \omega ,0,\tau _{1}\right]$). Later we investigate the role of nonzero $\sigma$ and the advantage of
introducing a finite time delay on cavity 1 output.

We now derive a simple analytic characterization of the output state at $%
\omega \simeq 0$ (i.e.~output light near the cavity resonances) which yields
insight into the generation of entanglement. 
We find that the state has the form of a 2-mode squeezed thermal state: 
\begin{equation}
\hat{\rho}_{12}=\hat{S}_{12}\left( R_{12}\right) \left[ \hat{\rho}_{1}^{%
\mathrm{th}}\left( \bar{n}_{1}\right) \otimes \hat{\rho}_{2}^{\mathrm{th}%
}\left( \bar{n}_{2}\right) \right] \hat{S}_{12}^{\dag }\left( R_{12}\right) .
\label{2modes}
\end{equation}%
Here $\hat{S}_{12}\left( R_{12}\right) =\exp \left[ R_{12}\hat{D}_{1}^{%
\mathrm{out}}[0]\hat{D}_{2}^{\mathrm{out}}[0]-h.c.\right] $ is the two-mode
squeeze operator, with $R_{12}$ the squeezing parameter, and $\hat{\rho}%
_{j}^{\mathrm{th}}\left( \bar{n}_{j}\right) $ describes a single-mode
thermal state with average population $\bar{n}_{j}$. The output state is
thus completely characterized by just 3 parameters: $\bar{n}_{1}$, $\bar{n}%
_{2}$ and $R_{12}$. In our case, they depend only on cavity cooperativities $%
C_{i}=4G_{i}^{2}/\left( \gamma \kappa _{i}\right) $ and bath temperatures.
The entanglement of a such a state is found to be~\cite{EPAPS}: 
\begin{equation}
E_{N}^{\mathrm{out}}[0]=-\ln \left( n_{R}-\sqrt{n_{R}^{2}-(1+2\bar{n}%
_{1})(1+2\bar{n}_{2})}\right) ,  \label{en2mthermal}
\end{equation}%
with $n_{R}=\left( \bar{n}_{1}+\bar{n}_{2}+1\right) \cosh 2R_{12}$. 

Assuming first the ideal situation where the cavity and mechanical baths are
at zero temperature, we find: 
\begin{eqnarray}
&&\bar{n}_{1}=0,\qquad \bar{n}_{2}=4C_{2}\frac{\gamma ^{2}}{\gamma _{\mathrm{%
\ tot}}^{2}}=\frac{4C_{2}}{\left( 1+C_{1}-C_{2}\right) ^{2}},  \label{n2} \\
&&\tanh 2R_{12}=\frac{4\sqrt{C_{1}C_{2}}\left( C_{2}+C_{1}+1\right) }{
C_{1}^{2}+C_{2}^{2}+6C_{1}C_{2}+2\left( C_{1}+C_{2}\right) +1}.\qquad
\label{R}
\end{eqnarray}
We stress that the effective thermal occupancies $\bar{n}_{j}$ are not
simply equal to bath temperatures; in particular, $\bar{n}_{2} \neq 0$ even
when all input noises are vacuum.

Equations~(\ref{2modes})-(\ref{R}) give us a simple way to understand
entanglement generation. The ideal situation is when $R_{12}\gg 1$ and $%
\bar{n}_i \rightarrow 0$. From Eq.~(\ref{R}), we see that a large $R_{12}$
generically requires large $C_1,C_2$. However, Eq.~(\ref{n2}) indicates that
this limit also yields a large effective temperature for cavity $2$
(heuristically, $\hat{H}_{\mathrm{int}}$ turns vacuum noise into thermal
noise). This heating degrades the purity of the state; however, as $\bar{n}_1 $ remains zero, 
it only slightly degrades the entanglement \cite{Wang2013}.

From Eqs.~(\ref{en2mthermal})-(\ref{R}), the zero-temperature output
entanglement is 
\begin{equation}
E_{N}^{\mathrm{out}}[0]=\ln \left( \frac{\left( 1+C_{1}-C_{2}\right) ^{2}}{%
A+B+2C_{2}\left( 1+2C_{1}\right) -4\sqrt{AB}}\right) ,
\label{ent_cavity_output}
\end{equation}%
with $A=C_{2}\left( C_{1}+C_{2}\right) $, $B=\left( 1+C_{1}\right)
^{2}+C_{1}C_{2}$. The approach outlined here can be easily extended to
finite temperature (see~\cite{EPAPS} for discussion). The results at both
zero and finite temperature are plotted in Fig.~\ref{fig:inputvsoutput}.

\begin{figure}
\begin{center}
\includegraphics[width=0.47\textwidth]{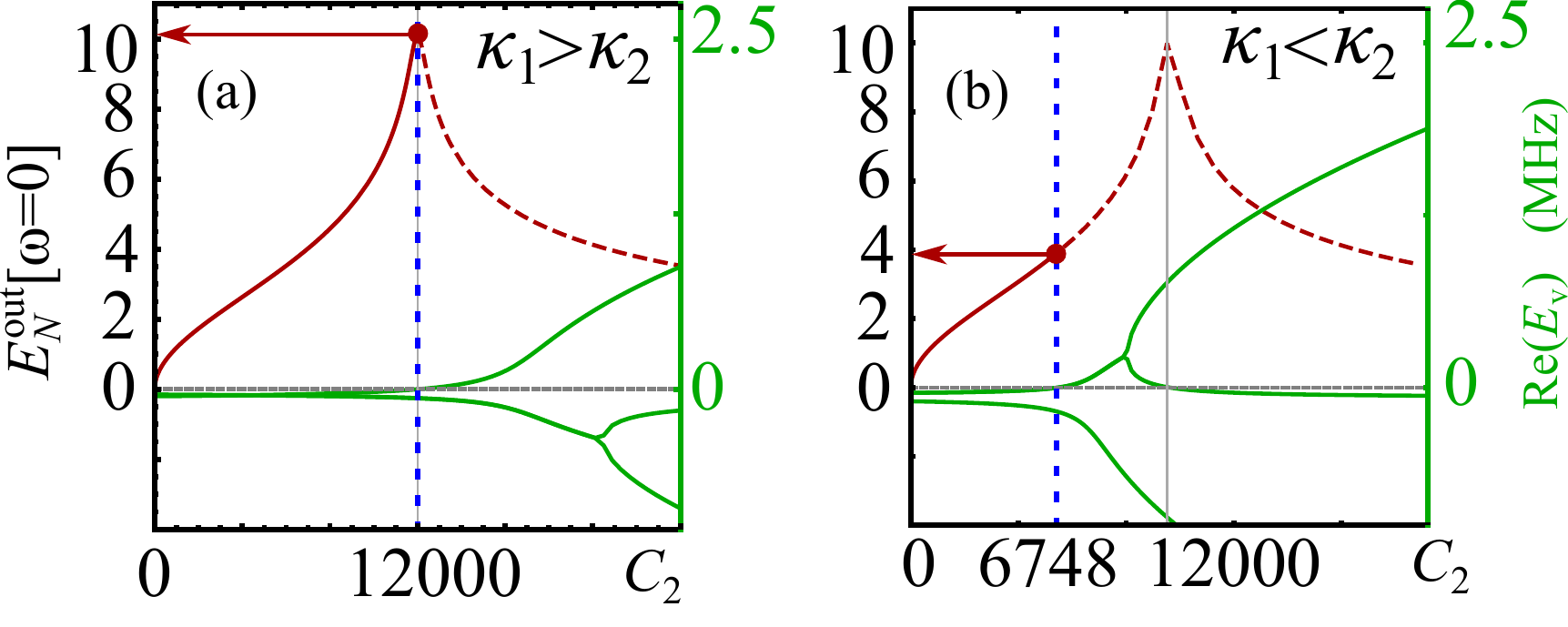}
\end{center}
\caption{Output entanglement of the two cavities fields at resonance (i.e.~$\omega=0 $) as a 
function of $C_2$ for $C_1=12000$. In (a), $\kappa_1>\kappa_2$ ($\kappa_1= 1.5 \kappa_2 = 2 
\pi \times 0.3$~MHz), in (b) $\kappa_1<\kappa_2$ ($\kappa_1=0.75 \kappa_2 = 2 \pi \times 0.3$~MHz).
Without considering the instability condition, the entanglement would reach
a maximum finite value at $C_2=C_1+1$ (gray vertical lines). The green lines
show the real part of the three eigenvalues ($E_\mathrm{v}$) of the
susceptibility matrix; the system becomes unstable when one of the real
parts becomes positive, which is indicated by the blue dashed lines. The red
dots indicates the maximum entanglement in each case given the constraints
of stability. We used $\gamma =2 \pi \times 10$~KHz. 
The red dashed line indicates the system has entered the unstable regime.
}
\label{fig:instability}
\end{figure}
\textit{Optimal output entanglement--} Eq.~(\ref{ent_cavity_output}) shows
that the zero-temperature entanglement is only a function of $C_{1}$, $C_{2}$, thus weak-coupling 
does not prevent strong entanglement; 
for fixed $C_1$, the expression is optimized when 
\begin{equation}
C_{2}=C_{1}+1 \equiv C_{2,\mathrm{opt}}.  \label{cond}
\end{equation}
Heuristically, this condition corresponds to having a total mechanical
damping $\gamma_{\mathrm{tot}} = 0$. While $E_{N}^{\mathrm{out}}[0]$ appears
to be only a function of the $C_i$, the ratio $\kappa_1 / \kappa_2$ also
plays an independent role via the stability condition of Eq.~(\ref{stability}%
). If $\kappa_1 \geq \kappa_2$, $C_{2,\mathrm{opt}}$ is also the maximum
value of $C_2$ for which the system is stable. In contrast, if $\kappa_2
\geq \kappa_1$, the system becomes unstable before $C_2$ reaches $C_{2,\mathrm{opt}}$,
hence one cannot achieve the optimal amount of entanglement. We thus see
that in addition to achieving large $C_i$ it is advantageous to have $%
\kappa_1 \geq \kappa_2$. Fig.~\ref{fig:instability} illustrates the
behavior of $E_{N}^{\mathrm{out}}[0]$ in these two cases ($C_{2,\mathrm{opt}%
}$ is indicated as a gray vertical line). Note that a similar dependence on $%
\kappa_1 / \kappa_2$ was observed numerically in~\cite{Tian2013},
but in a regime far from optimal entanglement (i.e.~for $C_2 \ll C_{2,%
\mathrm{opt}} $). 

In the large $C_{1}$ limit, the maximum achievable entanglement for the two
cases reduces to: 
\begin{equation}
E_{N}^{\mathrm{out}}[0]\Big |_{\mathrm{max}}\simeq \left\{ 
\begin{array}{ll}
\ln \left( \frac{2C_{1}}{1+2N_{\mathrm{m}}}\right) , & \text{\ if }\kappa
_{1}\geq \kappa _{2}, \\ 
-\ln \left[ \left( \frac{\kappa _{2}-\kappa _{1}}{\kappa _{2}+\kappa _{1}}%
\right) ^{2}+\frac{4\kappa _{2}^{2}\kappa _{1}N_{\mathrm{m}}^{\prime }}{%
C_{1}\left( \kappa _{2}+\kappa _{1}\right) ^{3}}\right] , & \text{\ if }%
\kappa _{1}<\kappa _{2},%
\end{array}%
\right.   \label{maxen}
\end{equation}%
with $N_{\mathrm{m}}^{\prime }=N_{\mathrm{m}}\left( 1+\kappa _{2}/\kappa
_{1}\right) +1$ and $N_{\mathrm{m}}$ the mechanical bath thermal occupancy.
In both cases, the output entanglement $E_{N}^{\mathrm{out}}$ is maximal at
the boundary of system stability, similar to the behavior of a
non-degenerate parametric amplifier (NDPA)~\cite{WallsMilburn}. However,
unlike a NDPA, $E_{N}^{\mathrm{out}}$ does not diverge at this boundary. 
Further, while $E_{N}^{\mathrm{out}}$ is maximal at this boundary, the
intra-cavity entanglement is extremely sub-optimal at this point. For $\kappa _{1}\geq
\kappa _{2} \gg \gamma$ and $C_{2}\simeq C_{2,\mathrm{opt}}\gg 1$, one has extremely
large entanglement of the output fields while simultaneously having almost
zero entanglement of the intracavity fields (see
Fig.~\ref{fig:inputvsoutput}).  

\begin{figure}[tbp]
\begin{center}
\includegraphics[width=0.47\textwidth]{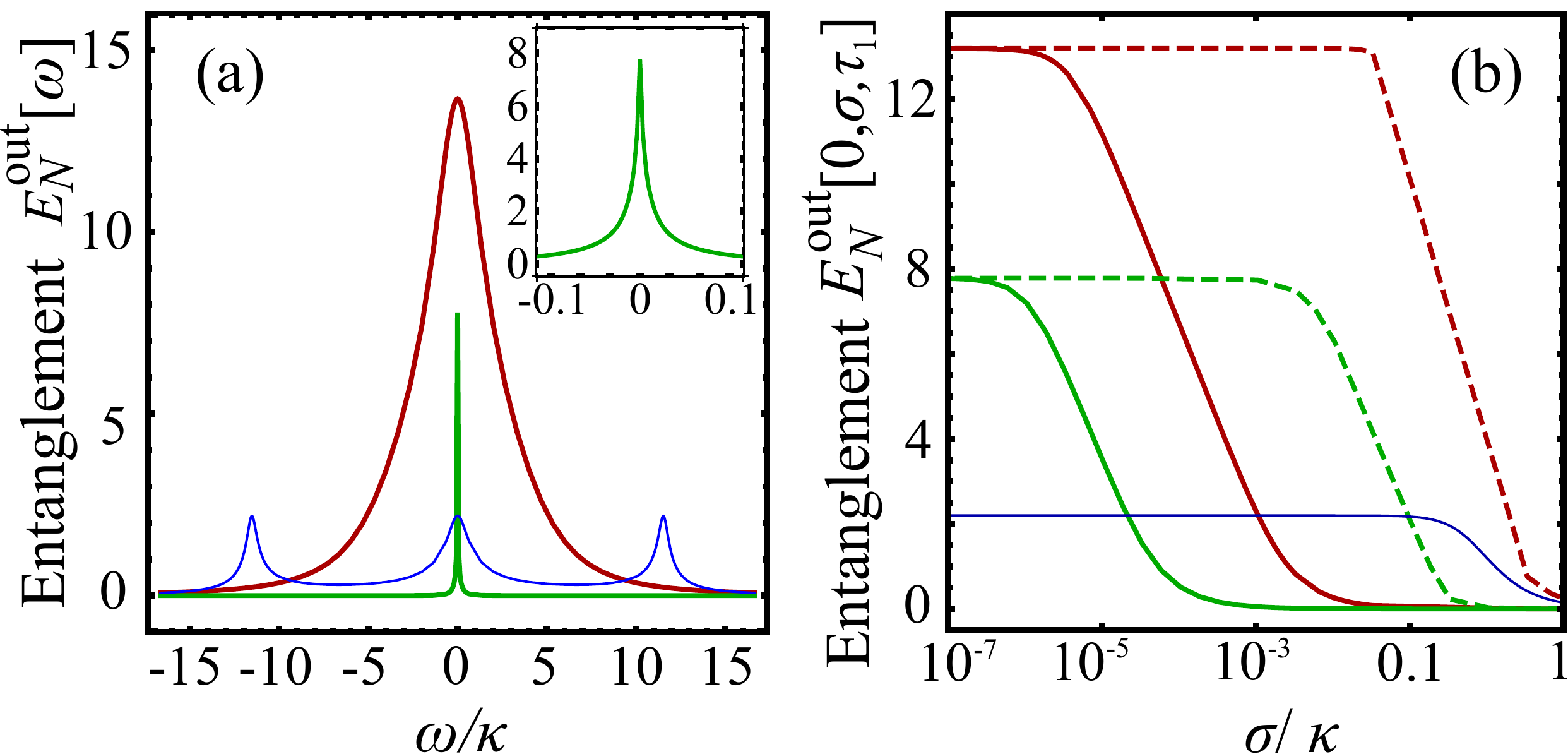}
\end{center}
\caption{ (a) Output entanglement in the limit of small mode bandwidth ($\sigma \rightarrow 0$) as a function of mode
center frequency $\omega$, for 3 cases: strong equal-coupling (red thick upper
line, $G_1 = G_2 = 13.3 \kappa$), weak equal-coupling (green thick lower line, $G_1 = G_2 = 0.1 \kappa$) and resolved normal-modes (blue thin line, $G_{1}/\kappa=13.3$ and $G_{2}/\kappa=6.7$. For the  the blue and red curves, $\gamma/\kappa =1.67\times 10^{-3}$ while for the green curve $\gamma/\kappa =3.3\times 10^{-5}$. The inset shows a zoom-in of the green curve (weak coupling case).
(b) Effects of non-zero mode bandwidth $\sigma$ on the output entanglement of modes with center frequency $\omega = 0$; colors
correspond to same parameters as in (a). The solid lines are for  zero time-delay between the two cavity output modes ($\tau_1=0$, c.f.~Eq.~(\ref{eq:DMode})) while the dashed
lines are the result including an optimal time delay $\tau_1=\kappa/(4G^2)$.}
\label{fig:1peakvs3peak}
\end{figure}

\textit{Frequency dependence-- } We now specialize to the case $\kappa
_{1}=\kappa _{2}=\kappa$, $C_{j}\gg 1$, but vary the output mode center frequency $%
\omega $. We consider two generic regimes. The first is that of equal
couplings $G_{1}=G_{2}=G$ (i.e., $C_{1}=C_{2}$) which, as discussed, is essentially optimal
for maximizing $E_{N}^{\mathrm{out}}[0]$ (it is also an ideal point to
generate quantum-limited amplification \cite{Metelmann2014}). 
In this regime, the effective coupling $\tilde{G}$ in Eq.~(%
\ref{eq:Hint}) vanishes, meaning that the three normal modes of $\hat{H}_{%
\mathrm{int}}$ are degenerate. Consequently, the $\sigma=0$ entanglement
spectrum $E_{N}^{\mathrm{out}}[\omega]$ has a single peak at $\omega =0$ (see thick curves in Fig.~\ref{fig:1peakvs3peak}) of
width $\sim \gamma C^{3/4}$ in the weak-coupling case and $\sim \sqrt{G}\left( 2\kappa ^{5}\gamma \right) ^{1/12} $ in the strong-coupling case. 
$E_{N}^{\mathrm{out}}[0,\sigma,\tau_1]$ decays on a similar scale as a function of mode bandwidth $\sigma$, if one
appropriately optimizes the time delay $\tau_1\sim \kappa/(4G^2)$; without including this delay, the decay
with $\sigma$ is much more pronounced (see Fig.~\ref{fig:1peakvs3peak}b). 
We stress that achieving large optimal $E_N$ in this regime only
requires strong cooperativities, and not the more stringent strong coupling
condition $G_{j}>\kappa _{j}$ (c.f. Eq.~(\ref{maxen})).

Keeping $\kappa_1 = \kappa_2$ and $C_j \gg 1$, another generic regime is
where $G_2 / G_1$ is sufficiently smaller than $1$ such that the effective
coupling $\tilde{G}$ is larger than $\kappa, \gamma$; this necessarily requires $G_1 > \kappa$.
 In this regime, the
three normal modes of $\hat{H}_{\mathrm{int}}$ are spectrally resolved and $%
E_{N}^{\mathrm{out }}[\omega]$ has correspondingly three peaks \cite%
{Tian2013}. The entanglement at $\omega = 0$ is necessarily much smaller
than the optimal value in Eq.~(\ref{maxen}).  One finds that $C_1 \gg 1$ is not
by itself enough to ensure large $E_N$ in this regime; one also needs to be deep
in the strong coupling regime, $G_1 \gg \kappa$.    
As discussed in~\cite{Tian2013}, this ``resolved-modes" regime does however offer enhanced protection against mechanical thermal
noise, as the central peak is due to the mechanically-dark normal mode $%
\beta_B$. 

Finally we note that, in contrast to the intracavity case~\cite{Wang2013},
output entanglement generation is extremely sensitive to any internal cavity loss $\kappa'$ and $E_N[0]$ is bounded by 
$\ln (\kappa_\mathrm{tot} / \kappa')$~\cite{EPAPS}, where $\kappa_\mathrm{tot} = \kappa + \kappa'$.

\textit{Hybrid 3-mode entangled state --} 
As discussed in the introduction, experimental setups using optomechanical
crystals~\cite{Eichenfield2009b,Safavi2011} could access the mechanical output
field via engineered phonon waveguides. Alternatively, the mechanical output can be
accessed by having mechanical dissipation dominated by an auxiliary cooling
cavity (Fig.~\ref{fig:inputvsoutput}); see \cite{EPAPS} for details.
Motivated by these possibilities, we now consider the multi-partite
entanglement properties of all three output fields in our system. Note that
previous work studied the non-stationary tripartite entanglement of
intracavity fields generated by the \textit{closed-system} Hamiltonian $\hat{%
H}_{\mathrm{int}}$ in Eq.~(\ref{eq:Hint}) \cite{Ferraro2004}; in contrast,
our focus here is on the steady-state output entanglement in our \textit{%
dissipative} 3-mode system.

We focus on zero frequency and zero bandwidth, and consider the ideal case
where all dissipative baths are at zero temperature. In this case, the
3-mode output state is pure, and can be written as a twice-squeezed vacuum~%
\cite{EPAPS}: 
\begin{equation}
\left\vert \Psi \right\rangle =\hat{S}_{12}\left( R_{12}\right) \hat{S}_{2%
\mathrm{m}}\left( R_{2\mathrm{m}}\right) \left\vert 000\right\rangle ,
\label{3modes}
\end{equation}%
where $\hat{S}_{2\mathrm{m}}\left( R_{2\mathrm{m}}\right) \equiv \exp \left[
iR_{2\mathrm{m}}\hat{D}_{\mathrm{m}}^{\mathrm{out}}\left[ 0\right] \hat{D}%
_{2}^{\mathrm{out}}[0]-h.c.\right] $ is the 2-mode squeeze operator
entangling the output of the mechanics at $\omega =0$ (denoted by $\hat{D}_{%
\mathrm{m}}^{\mathrm{out}}\left[ 0\right] $, which is defined in a similar
way as the cavity output) with that of cavity 2. The squeezing parameter is $%
R_{2\mathrm{m}}=\sinh ^{-1}\bar{n}_{2}$, where $\bar{n}_{2}$ is given in
Eq.~(\ref{n2}) and diverges at the instability point. Equation~(\ref{3modes}) demonstrates that the effective
temperature $\bar{n}_{2}$ which degraded the cavity-cavity entanglement in
Eq.~(\ref{en2mthermal}) is a direct consequence of entanglement between
cavity $2$ and the mechanics. It also demonstrates the asymmetry between
the three modes (i.e.,~there is no direct squeezing between the mechanical and cavity 1 outputs.)

More strikingly, Eq.~(\ref{3modes}) shows that we have genuine tripartite
entanglement (GTE): none of the parties can be separated from any other
in a mixture of product states, implying a fully inseparable state \cite{Giedke2001,vanLoock2000}. 
To see this, note that the total state
is pure, and if one traces over one subsystem, the remaining two are in a
mixed state. GTE is also evident by writing the state in the Fock-state
basis $|n_{1},n_{2},n_{\mathrm{m}}\rangle $,%
\begin{equation}
\left\vert \Psi \right\rangle =\sum_{pq}\frac{\sqrt{C_{p+q}^{p}}\left\langle
N_{\mathrm{m}}\right\rangle ^{\frac{q}{2}}\left\langle N_{1}\right\rangle ^{%
\frac{p}{2}}}{\left( 1+\left\langle N_{2}\right\rangle \right) ^{(p+q+1)/2}}%
\left\vert p,p+q,q\right\rangle ,  \label{expand}
\end{equation}%
where $C_{p+q}^{p}$ are binomial coefficients and $\left\langle
N_{i}\right\rangle =\left\langle \left( \hat{D}_{i}^{\mathrm{out}}[0]\right)
^{\dag }\hat{D}_{i}^{\mathrm{out}}[0]\right\rangle $ ($i=1,2,\mathrm{m}$) is
the average photon/phonon number of each mode (see \cite{EPAPS} for their
values). $\left\vert \Psi \right\rangle $ only involves Fock states $%
|n_{1},n_{2},n_{\mathrm{m}}\rangle $ where $n_{2}=n_{1}+n_{\mathrm{m}}$;
there is thus a perfect correlation between the three systems that is only
evident by looking at all three modes. If any two modes are traced out, the
remaining third mode is in an impure thermal state. Notice that, in the absence of dissipation, 
a similar correlated state of intra-cavity quanta can be generated by
the Hamiltonian in Eq.~(\ref{eq:Hint})~\cite{EPAPS,Ferraro2004}.

GTE can also be
quantified using the Gaussian R\'enyi-2 measure recently introduced in~\cite{Adesso2012}. We find that it is indeed non-zero, and diverges as
one increases $C_{2}$ to the instability point $C_{1}+1$ (due to the divergence of $\bar{n}_2$ and $R_{2\mathrm{m}}$)
\cite{EPAPS}. These calculations also
reveal the absence of any direct entanglement between the mechanical and cavity 1 outputs. 

\textit{Conclusion --} We have studied the bipartite and
tripartite entanglement of the output fields in a 3-mode optomechanical
system. For bipartite photonic entanglement, an explicit analytical expression
enables us to obtain an impedance matching
condition which yields maximal entanglement. We also have shown that
the three output fields are in a genuine tripartite entangled state, with entanglement diverging
near the boundary of system stability.

We thank M. Woolley for useful discussions. This work was supported by the
DARPA ORCHID program under a grant from the AFOSR. SC and YDW acknowledge support from Chinese Youth 1000 Talents Program.

Note added.- During the preparation of this work, we came to learn of a related work by Deng, Habraken, 
and Marquardt looking at different aspects of output light entanglement in an optomechanical system.

\begin{widetext}

\newpage

\renewcommand{\theequation}{S\arabic{equation}}

\setcounter{equation}{0}

\section{Supplemental information}

\section{Scattering matrix}

The scattering matrix $\mathbf{S}[\omega ]$ is obtained via the
system Langevin equations (with RWA) and input-output relations. In our
interaction picture:
\begin{equation}
\mathbf{S}[ \omega ] =\mathbf{1}+\frac{1}{C_{1}\chi _{1}\chi _{%
\mathrm{m}}-C_{2}\chi _{2}\chi _{\mathrm{m}}+1}\left( 
\begin{array}{ccc}
\chi _{1}\left( \frac{C_{2}}{4}\chi _{2}\chi _{\mathrm{m}}-1\right) & \frac{%
\sqrt{C_{1}C_{2}}}{4}\chi _{1}\chi _{2}\chi _{\mathrm{m}} & \frac{iC_{1}}{2}%
\chi _{1}\chi _{\mathrm{m}} \\ 
-\frac{\sqrt{C_{1}C_{2}}}{4}\chi _{1}\chi _{2}\chi _{\mathrm{m}} & -\chi
_{2}\left( \frac{C_{1}}{4}\chi _{1}\chi _{\mathrm{m}}+1\right) & -\frac{%
iC_{2}}{2}\chi _{2}\chi _{\mathrm{m}} \\ 
\frac{iC_{1}}{2}\chi _{1}\chi _{\mathrm{m}} & \frac{iC_{2}}{2}\chi _{2}\chi
_{\mathrm{m}} & -\chi _{\mathrm{m}}%
\end{array}%
\right) ,
\end{equation}%
with $\chi _{i}=2\kappa _{i}/(\kappa _{i}-2i\omega )$ ($i\in \{1,2,m\}$, $%
\kappa _{\mathrm{m}}\equiv \gamma $). Notice that at $\omega=0$, $\chi _{i}=2$, the matrix 
only depends on the cooperativities.

\section{Bipartite entanglement at finite temperature}

\subsection{Mapping to a two-mode squeezed thermal state}

Using the standard approach (i.e. plugging the system covariance matrix into
the definition of logarithmic negativity), the entanglement of the 3-mode
system can be easily computed numerically. However, we find that mapping the
output state to a 2-mode squeezed thermal state can yield a simple
analytical expression which reveals a number of intresting properties of the
output entanglement.

It can be seen from the Langevin equation of the system that $\left\langle \hat{D}_{i}^{\mathrm{out}}%
\left[ \omega \right] \right\rangle $ are all zero and only 3
correlators of the output cavity modes are nonzero: $\left\langle \left( 
\hat{D}_{1}^{\mathrm{out}}\left[ \omega \right] \right) ^{\dag }\hat{D}_{1}^{%
\mathrm{out}}\left[ \omega \right] \right\rangle $, $\left\langle \left( 
\hat{D}_{2}^{\mathrm{out}}\left[ \omega \right] \right) ^{\dag }\hat{D}_{2}^{%
\mathrm{out}}\left[ \omega \right] \right\rangle $ and $\left\langle \hat{D}%
_{1}^{\mathrm{out}}\left[ \omega \right] \hat{D}_{2}^{\mathrm{out}}\left[
\omega \right] \right\rangle $. A two-mode squeezed
thermal state has the same covariance matrix. Since two Gaussian states with
the same covariance matrix represent the same state, this output cavity
state can be mapped to a two-mode squeezed thermal state whose non-zero
correlation function are simply%
\begin{eqnarray}
\left\langle \left( \hat{D}_{i}^{\mathrm{out}}\left[ \omega \right] \right)
^{\dag }\hat{D}_{i}^{\mathrm{out}}\left[ \omega \right] \right\rangle &=&%
\bar{n}_{i}+\left( \bar{n}_{1}+\bar{n}_{2}+1\right) \sinh ^{2}\left\vert
R_{12}\left[ \omega \right] \right\vert ,  \notag \\
\left\langle \hat{D}_{1}^{\mathrm{out}}\left[ \omega \right] \hat{D}_{2}^{%
\mathrm{out}}\left[ \omega \right] \right\rangle &=&-e^{i\theta \left[
\omega \right] }\left( \bar{n}_{1}+\bar{n}_{2}+1\right) \sinh \left\vert
R_{12}\left[ \omega \right] \right\vert \cosh \left\vert R_{12}\left[ \omega %
\right] \right\vert ,  \label{DD_correl}
\end{eqnarray}%
where $\bar{n}_{i}\equiv \bar{n}_{i}\left[ \omega \right] $, $R_{12}\left[
\omega \right] $ is complex in general and $\theta \left[ \omega \right]
\equiv \arg \left( R_{12}\left[ \omega \right] \right) $. Plugging in the
correlators with $\sigma =0$ and $\omega =0$, we find $R_{12}$ is real and
the output state can be characterized with the 3 parameters%
\begin{eqnarray}
\bar{n}_{1} &=&\frac{1}{2}\left( \frac{4N_{\mathrm{m}}}{C_{1}-C_{2}+1}-\frac{%
4\left( C_{2}+N_{\mathrm{m}}\right) }{\left( C_{1}-C_{2}+1\right) ^{2}}+%
\frac{\sqrt{E^{2}-D^{2}}}{\left( C_{1}-C_{2}+1\right) ^{2}}-1\right) , 
\notag \\
\bar{n}_{2} &=&\frac{1}{2}\left( \frac{-4N_{\mathrm{m}}}{C_{1}-C_{2}+1}+%
\frac{4\left( C_{2}+N_{\mathrm{m}}\right) }{\left( C_{1}-C_{2}+1\right) ^{2}}%
+\frac{\sqrt{E^{2}-D^{2}}}{\left( C_{1}-C_{2}+1\right) ^{2}}-1\right) , 
\notag \\
\tanh 2R_{12} &=&D/E,  \label{n1n2T}
\end{eqnarray}%
with%
\begin{eqnarray}
D &=&4\sqrt{C_{1}C_{2}}\left( C_{1}+C_{2}+1+2N_{\mathrm{m}}\right) ,
\label{n1n2Taux} \\
E &=&\left( C_{1}+C_{2}\right) ^{2}+2\left( C_{1}+C_{2}\right) \left( 1+2N_{%
\mathrm{m}}\right) +1+4C_{1}C_{2}.
\end{eqnarray}%
Notice that such a mapping is unique, i.e., there isn't an alternate choice
of $\bar{n}_{1}$, $\bar{n}_{2}$ and $R_{12}$ that yields the same covariance
matrix. At zero temperature, the parameters are reduced to Eqs.~(6)-(7) of
the main text.

Plugging Eq.~(\ref{n1n2T}) into Eq.~(5) of the main text, one can obtain the
general result of entanglement at finite temperature. Up to linear order in $%
N_{\mathrm{m}}$, it reads%
\begin{equation}
E_{N}^{\mathrm{out}}\left[ 0\right] \approx E_{N}^{\mathrm{out}\text{,}(0)}%
\left[ 0\right] +\left( C_{1}+C_{2}-\sqrt{\frac{\left( 1+2C_{1}\right)
^{2}C_{2}\left( C_{1}+C_{2}\right) }{1+C_{1}^{2}+C_{1}\left( 2+C_{2}\right) }%
}\right) \frac{4N_{\mathrm{m}}}{\left( C_{1}-C_{2}+1\right) ^{2}}+O(N_{%
\mathrm{m}}^{2}),
\end{equation}%
where $E_{N}^{\mathrm{out}\text{,}(0)}\left[ 0\right] $ is the entanglement
at zero temperature as shown in Eq.~(8) of the main text.

\subsection{Entanglement spectrum}

As discussed in the main text, the bipartite entanglement of the cavity
output modes can be quantified using the logarithmic negativity and for now
we discuss the spectrum of entanglement without considering finite
bandwidth, i.e. $E_{N}^{\mathrm{out}}\left[ \omega \right] \equiv E_{N}^{%
\mathrm{out}}\left[ \omega ,0,\tau _{1}\right] $. When $G_{1}=G_{2}$ and $%
\kappa _{1}=\kappa _{2}$, there is only one peak centered at $\omega =0$ in
the spectrum of the output entanglement as shown by the red and green thick
curves in Fig. \ref{fig:1peakvs3peaksupp} (a) (a re-plot of Fig. 3(a) of the
main text). The peak value at $\omega =0$ is independent of strong-coupling
condition%
\begin{equation}
E_{N}^{\mathrm{out}}\left[ 0\right] \approx \ln \left( \frac{2C}{1+2N_{%
\mathrm{m}}}\right) +\mathcal{O}\left( C^{-1}\right) ,  \label{enteq1}
\end{equation}%
In the regime of sufficiently strong coupling, we find a simple
expression for the half width at half maximum (HFHM) of this peak: $\Delta \omega \approx 
\sqrt{G}\left( 2\kappa ^{5}\gamma \right) ^{1/12}$ under the condition that $%
\left( G/\kappa \right) ^{6}>\kappa /\gamma \gg 1$ and $C\gg 1$. We stress
that a robust peak in $E_{N}^{\mathrm{out}}\left[ \omega \right] $ remains
in the weak-coupling case $G<\kappa $ as long as $C\gg 1$; the HWHM is $\Delta \omega \approx \gamma (2C)^{3/4}$.

The entanglement at $\omega =0$ in the equal-coupling regime shows a
prominent decrease with temperature due to the mechanical thermal noise. At
low temperature, the equal-coupilngs ($G_{1}=G_{2}$) regime yields a decent
entanglement even for weak coupling ($\kappa >G$) as shown by the green
curve in Fig. \ref{fig:1peakvs3peaksupp} (b).

\begin{figure}[tbp]
\begin{center}
\includegraphics[width=0.8\textwidth]{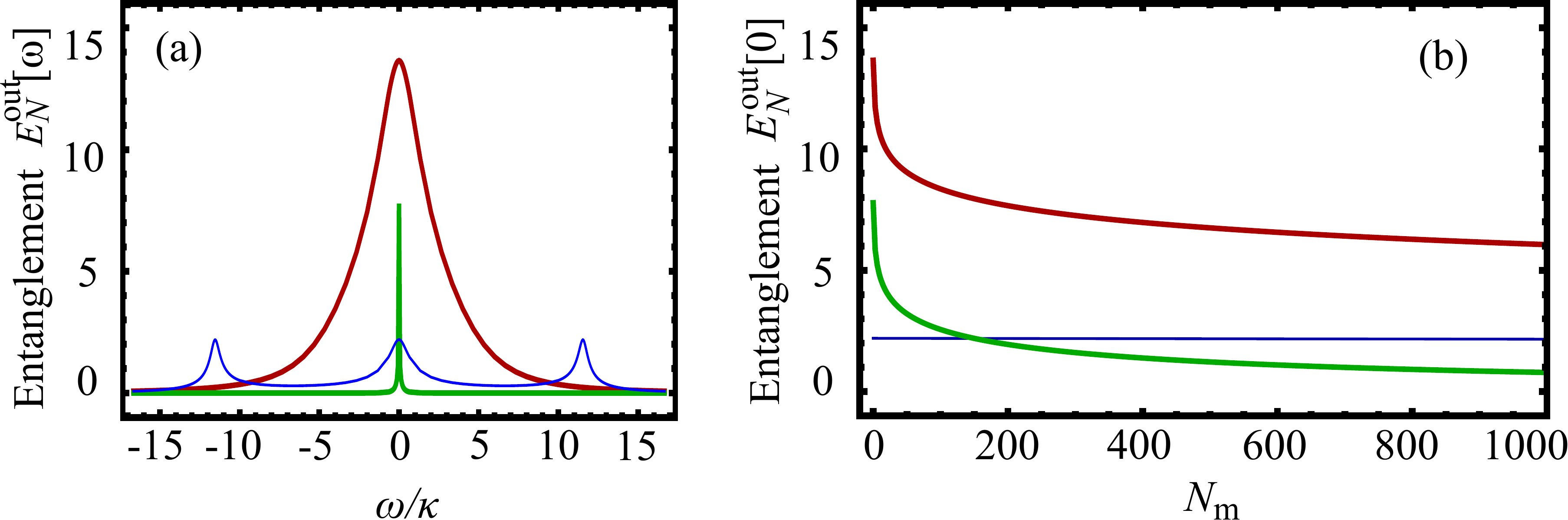}
\end{center}
\caption{Output entanglement for 3 cases: strong equal couplings (thick red
(upper) lines, $G/\kappa =13.3$), weak equal couplings (thick green
(lower) lines, $G/\kappa =0.1$) and resolved normal modes (blue thin
lines, $G_{1}/\kappa =13.3$ and $G_{2}/\protect\kappa =6.7$). Panel
(a) is a re-plot of Fig.~3(a) in the main text, showing the spectrum of
output entanglement with $N_{\mathrm{m}}=0$. Panel (b) shows the output
entanglement at $\omega =0$ \emph{vs.} mechanical temperature $N_{m}$. 
For the green curves, $\gamma /\kappa =3.3\times 10^{-5}$; and for the rest, $\gamma /\kappa =1.67\times 10^{-3}$.}
\label{fig:1peakvs3peaksupp}
\end{figure}

We turn now to the regime where $\tilde{G}=\sqrt{G_{1}^{2}-G_{2}^{2}}>\kappa 
$, where the three normal modes of the interaction Hamiltonian in Eq.~(1) of
the main text are spectrally resolved. In this regime (as discussed in \cite%
{Tian2013supp}), $E_{N}^{\mathrm{out}}\left[ \omega \right] $ has three peaks,
with each peak corresponding to a normal mode. The \textquotedblleft dark
mode\textquotedblright\ $\beta _{B}$ corresponds to the peak at $\omega =0$,
while the peaks at $\pm \tilde{G}$ are the coupled \textquotedblleft
hybrid\textquotedblright\ modes (see blue thin curve in Fig. \ref%
{fig:1peakvs3peaksupp} (a)); all have width $\sim \kappa $. In the simple case $%
\kappa _{1}=\kappa _{2}$, resolving the normal modes requires $G_{1}>\kappa
\cosh r\gg \kappa $, i.e., deep in the strong coupling regime for a large $r$%
. For typical parameters, the maximum entanglement in the resolved-peaks
regime is far less that the optimal value that would be achieved if $C_{2}$
were increased to $C_{1}$ (see Fig. \ref{fig:1peakvs3peaksupp} (b)). On the
other hand, the output entanglement at the central peak can be written as
(assuming $N_{\mathrm{m}}e^{2r}\ll \tilde{C}$) 
\begin{equation}
E_{N}^{\mathrm{out}}\left[ 0\right] \approx 4r-2e^{2r}\left( 2N_{\mathrm{m}%
}+1\right) /\tilde{C},
\end{equation}%
with the effective cooperativity $\tilde{C}=4\tilde{G}^{2}/\gamma \kappa $.
This shows that, as the central peak corresponds to the dark mode, the
mechanical noise is suppressed by the effective cooperativity and the
entanglement is robust to thermal fluctuation as shown in Fig.~\ref%
{fig:1peakvs3peaksupp} (b). The entanglement at the side peaks is still
sensitive to thermal noise and in the large $r$ limit: 
\begin{equation}
E_{N}^{\mathrm{out}}\left[ \omega =\pm \tilde{G}\right] \approx 2r-\ln
\left( \frac{4\gamma }{\kappa }\left( N_{\mathrm{m}}+\frac{1}{2}\right)
\right) ,
\end{equation}%
valid if $\gamma N_{\mathrm{m}}/\kappa \ll 1$.

\section{Influence of cavity internal losses on the bipartite entanglement}

In contrast to the generation of intracavity entanglement~\cite{Wang2013supp},
the output entanglement is sensitive to the internal losses of the cavities.
With internal loss, the total cavity damping rate becomes $\kappa
_{tot}=\kappa +\kappa ^{\prime }$, where $\kappa ^{\prime }$ describes
internal loss, and $\kappa $ is associated with the coupling to the output
field. We discuss the influence of internal loss in the following two
limiting cases.

1) Equal coupling case. With finite internal loss $\kappa _{1}^{\prime
}=\kappa _{2}^{\prime }=\kappa ^{\prime }$ and also assuming $G\gg \gamma $, 
$\kappa $, $\kappa ^{\prime }$, the entanglement reads 
\begin{equation}
E_{N}^{\mathrm{out}}\left[ 0\right] \approx -\ln \left( \frac{\kappa
^{\prime }}{\kappa ^{\prime }+\kappa }+\frac{1}{2C}\right) ,  \label{entkp}
\end{equation}%
which recovers the two limits: $E_{N}^{\mathrm{out}}\left[ 0\right] \approx
\ln 2C$ at large $\kappa $ (still, $C\gg 1$), and $E_{N}^{\mathrm{out}}\left[
0\right] \approx \ln \left( 1+\kappa /\kappa ^{\prime }\right) $ when $%
\kappa /\kappa ^{\prime }\ll 2C$. In the case of a cavity with tunable
external damping rate, the optimal $\kappa $ satisfies $\kappa =\kappa
^{\prime }\left( \sqrt{2C^{\prime }}-1\right) $. The corresponding
entanglement is $E_{N}^{\mathrm{out}}\left[ 0\right] \approx \frac{1}{2}\ln
(C^{\prime }/2)$, with $C^{\prime }=4G^{2}/\kappa ^{\prime }\gamma $. This
result is the same as the optimal intracavity entanglement, as shown in Eq.
(11) of \cite{Wang2013supp} (notice that the definition of $C$ in this paper
differs by a factor of $4$ comparing with \cite{Wang2013supp}). The maximum
entanglement for both spectral entanglement and intracavity entanglement are
the same.

2) Resolved peaks case. Including the internal loss and assuming $\kappa
_{i}^{\prime }=\kappa ^{\prime }$, $\tilde{G}\gg \kappa $, $\gamma $, $%
\kappa ^{\prime }$ 
\begin{equation}
E_{N}^{\mathrm{out}}\left[ 0\right] \approx 4r-\ln \left( \frac{\kappa
+e^{4r}\kappa ^{\prime }}{\kappa +\kappa ^{\prime }}+\frac{e^{r}\sinh r}{%
\tilde{C}}\right) .  \label{k1}
\end{equation}%
When $\kappa /\kappa ^{\prime }\ll e^{4r}\tilde{C}/(e^{2r}+\tilde{C})$, this
reduces to $E_{N}^{\mathrm{out}}\left[ 0\right] \approx \ln \left( 1+\kappa
/\kappa ^{\prime }\right) $. The maximum of entanglement in Eq. (\ref{k1})
is obtained when $\kappa _{\mathrm{opt}}=\kappa ^{\prime }\left( \sqrt{%
\tilde{C}\left( 1+e^{2r}\right) /2}-1\right) $ 
\begin{equation}
E_{N,\mathrm{opt}}^{\mathrm{out}}\left[ 0\right] =4r-\ln \left( 1+\sqrt{%
\frac{2}{\tilde{C}}}\frac{e^{4r}-1}{\sqrt{e^{2r}+1}}+\frac{1-e^{2r}}{\tilde{C%
}}\right) .
\end{equation}%
However, notice that in order to have resolved peaks, $\tilde{G}\gg \kappa
+\kappa ^{\prime }$. This means the optimal condition is normally not
satisfied unless $\kappa ^{\prime }\ll \gamma $.

\section{Influence of non-rotating wave terms}

In the main text, we only discussed the dynamics with the rotating wave
approximation (RWA), considering the good cavity limit $\kappa \ll \omega _{%
\mathrm{m}}$. Here we will give the full result including the non-RWA terms
and the precise condition to neglect them.

First we notice that the counter-rotating terms are time-independent in the
rotating frame with respect to the cavity drives. In this frame, the full
Hamiltonian is written as $\hat{H}=\omega _{\mathrm{m}}\left( \hat{b}^{\dag }%
\hat{b}+\hat{d}_{1}^{\dag }\hat{d}_{1}-\hat{d}_{2}^{\dag }\hat{d}_{2}\right)
+\hat{H}_{\mathrm{int}}+\hat{H}_{\mathrm{CR}}$ with%
\begin{eqnarray}
\hat{H}_{\mathrm{int}} &=&\left( G_{1}\hat{b}^{\dag }\hat{d}_{1}+G_{2}\hat{b}%
\hat{d}_{2}\right) +h.c.  \notag \\
\hat{H}_{\mathrm{CR}} &=&\left( G_{1}\hat{b}^{\dag }\hat{d}_{1}^{\dag }+G_{2}%
\hat{b}\hat{d}_{2}^{\dag }\right) +h.c..
\end{eqnarray}%
Thus a closed set of equations in the frequency domain can be obtained

\begin{eqnarray}
-i\left( \omega -\omega _\mathrm{m}\right) \hat{b}\left[ \omega \right] &=&-\frac{%
\gamma }{2}\hat{b}\left[ \omega \right] -i\left( G_{1}\left( \hat{d}_{1}%
\left[ \omega \right] +\hat{d}_{1}^{\dag }\left[ \omega \right] \right)
+G_{2}\left( \hat{d}_{2}^{\dag }\left[ \omega \right] +\hat{d}_{2}\left[
\omega \right] \right) \right) -\sqrt{\gamma }\hat{b}_{in}\left[ \omega %
\right]  \notag \\
-i\left( \omega -\omega _\mathrm{m}\right) \hat{d}_{1}\left[ \omega \right] &=&-%
\frac{\kappa _{1}}{2}\hat{d}_{1}\left[ \omega \right] -iG_{1}\left( \hat{b}%
\left[ \omega \right] +\hat{b}^{\dag }\left[ \omega \right] \right) -\sqrt{%
\kappa _{1}}\hat{d}_{in,1}\left[ \omega \right]  \notag \\
-i\left( \omega -\omega _\mathrm{m}\right) \hat{d}_{2}^{\dag }\left[ \omega \right]
&=&-\frac{\kappa _{2}}{2}\hat{d}_{2}^{\dag }\left[ \omega \right]
+iG_{2}\left( \hat{b}\left[ \omega \right] +\hat{b}^{\dag }\left[ \omega %
\right] \right) -\sqrt{\kappa _{2}}\hat{d}_{in,2}^{\dag }\left[ \omega %
\right]
\end{eqnarray}%
These equations can be solved analytically. Fig. \ref{fig:nonrwa} shows the
comparison of the results $E_{N}^{\mathrm{out}}\left[ \omega =0,\sigma =0%
\right] $ with/without RWA. The lower curve corresponds to the case where $%
\tilde{G}>\kappa $ and the normal modes are resolved: counter-rotating terms
suppress the entanglement, but become insignificant once $\omega _{\mathrm{m}%
}>\kappa $. The two upper curves correspond to the case with equal-coupling. In
the good cavity limit, the maximum entanglement in the equal-coupling regime
is much larger than the resolved-peak regime. Consequently, non-RWA
corrections play a larger role, and one can only achieve the RWA result deep
into the good cavity limit $\omega _{\mathrm{m}}\gg \kappa $.

For both equal-coupling and resolved-peak cases, the leading non-RWA
correction to $E_{N}^{\mathrm{out}}\left[ 0\right] $ is 
\begin{equation}
\delta E_{N}^{\mathrm{out}}\left[ 0\right] \approx -e^{E_{N}^{\mathrm{out}}%
\left[ 0\right] }\frac{\kappa ^{2}}{16\omega _{\mathrm{m}}^{2}}.
\label{den_nrwa}
\end{equation}%
Here $E_{N}^{\mathrm{out}}\left[ 0\right] $ is the entanglement with
rotating wave approximation. This approximate expression is shown by the blue
dashed line in Fig.~\ref{fig:nonrwa} (b). Thus the condition to
justify the use of RWA is%
\begin{equation}
\left\vert \frac{\omega _{\mathrm{m}}}{\kappa }\right\vert \gg \frac{1}{4}%
\sqrt{\frac{e^{E_{N}^{\mathrm{out}}}}{E_{N}^{\mathrm{out}}}},
\end{equation}%
which is looser in the case of resolved-peaks (due to the smaller $E_{N}^{%
\mathrm{out}}$).

\begin{figure}[tbp]
\begin{center}
\includegraphics[width=0.8\textwidth]{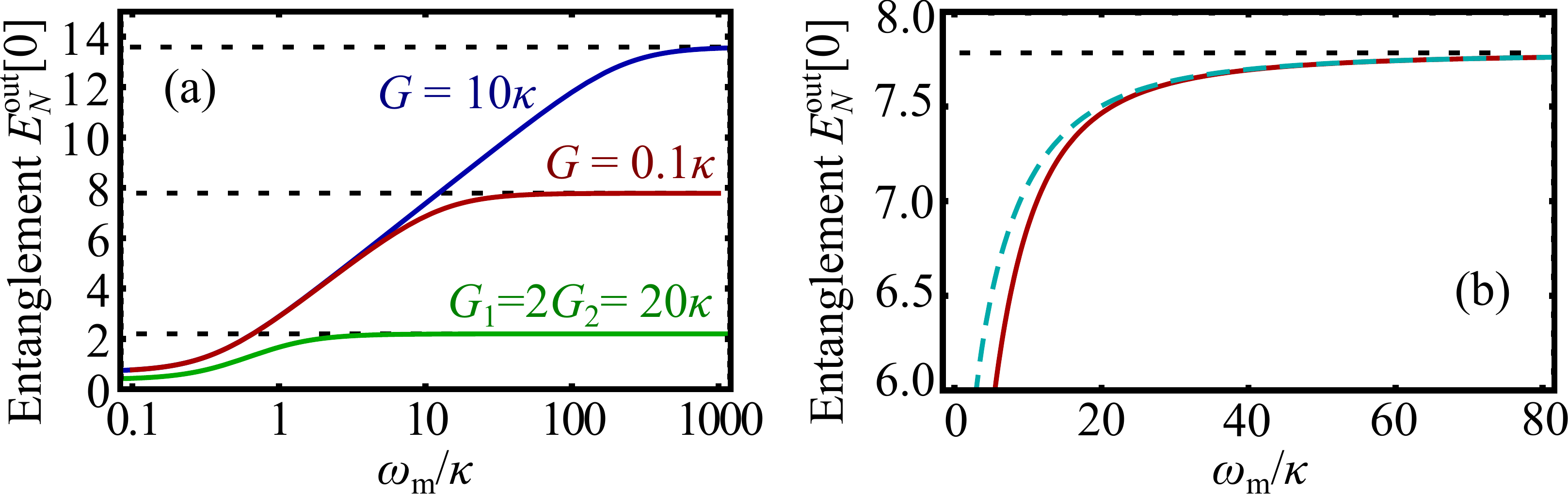}
\end{center}
\caption{Validity of the rotating-wave approximation. (a) the
solid lines show the output entanglement $E_N^\mathrm{out}[\omega=0]$
\emph{vs.} $\omega _{\mathrm{m}}/\protect\kappa $ including the
counter-rotating terms. From top to bottom, the 3 curves show the case of
strong equal-coupling ($G_1=G_2=10\kappa$), weak equal-coupling ($%
G_1=G_2=0.1\kappa$), resolved peak ($G_1=2G_2=10\kappa$).
The dashed lines mark the results with the rotating wave approximation. For
the upper (blue) and lower (green) curve, $\gamma /\kappa=1\times 10^{-3}$, while for the middle curve $\gamma /
\kappa =3.3\times 10^{-5}$. (b) shows the comparison of the
analytical expression Eq.~(\ref{den_nrwa}) (dashed line) with
the numerical result (solid line) in the weak coupling case ($G_1=G_2=0.1\kappa$). }
\label{fig:nonrwa}
\end{figure}

\section{Time delay to improve the entanglement bandwidth}

So far, we have considered the output entanglement with zero
bandwidth, i.e., $E_{N}^{\mathrm{out}}[\omega ,\sigma =0,\tau _{1}]$.
However, in practice, the filter function is of finite bandwidth. For
simplicity, we consider a square filter function centered at $\omega $ with
bandwidth $\sigma $, i.e.,%
\begin{equation}
\hat{D}_{i}^{\mathrm{out}}\left[ \omega ,\sigma ,\tau _{i}\right] =\frac{1}{%
\sqrt{\sigma }}\int_{\omega -\frac{\sigma }{2}}^{\omega +\frac{\sigma }{2}%
}d\omega ^{\prime }e^{-i\omega ^{\prime }\tau _{i}}\hat{d}_{i}^{\mathrm{out%
}}\left[ \omega ^{\prime }\right] .  \label{Didef}
\end{equation}%
Assuming the center frequency is set at $\omega =0$ (the cavity resonance
frequency in the lab frame), the entanglement has a non-trivial dependence
on the bandwidth $\sigma $ and the relative time delay $\tau _{1}$ (taking $%
\tau _{2}=0$).

The solid lines of Fig. 3(b) in the main text show the entanglement of the two
output cavity modes for zero time delay $E_{N}^{\mathrm{out}}[\omega
=0,\sigma ,\tau _{1}=0]$ as a function of bandwidth $\sigma $. While the
equal-coupling case yields large entanglement at $\sigma =0$, it is much
more sensitive to the increase of the bandwidth. $E_{N}^{\mathrm{out}%
}[0,\sigma,0]$ decays on a scale $\sigma \sim C^{-1/4}\gamma $. In
contrast, in the resolved normal-mode case, $E_{N}^{\mathrm{out}}[0,\sigma
,0]$ is less sensitive to increasing the mode bandwidth, and is only
suppressed significantly when $\sigma \sim \kappa $.

The strong sensitivity to non-zero $\sigma $ in the case of equal-coupling
is related to the change of squeezing phase at different $\omega $. As
discussed before, the cavity output state can be characterized by a 2-mode
squeezed thermal state (see Eq. (4) in the main text) with a complex
squeezing parameter $R_{12}\left[ \omega \right] $ whose phase is $\theta \left[
\omega \right] = \arg \left\langle -\hat{D}_{1}^{\mathrm{out}}\left[
\omega \right] \hat{D}_{2}^{\mathrm{out}}\left[ \omega \right] \right\rangle 
$ (Eq. (\ref{DD_correl})). At $\omega =0$ and $\sigma =0$, $R_{12}\left[ 0%
\right] $ is real (cf. Eq. (7) in the main text), i.e., $\theta =0$. For $%
\omega \neq 0$, a frequency-dependent phase arises (see Fig. \ref%
{fig:squeezingparameter}(b)). In the large squeezing limit, i.e., $%
|R_{12}|\gg 1$ (see Fig. \ref{fig:squeezingparameter}(a)), such a phase
variation leads to a rapid decrease of entanglement as $\sigma $ increases.
Assuming $\kappa _{1}=\kappa _{2}$ and $G_{1}=G_{2}$ where the magnitude of
squeezing is maximized, the peak width of $|R_{12}|$ is comparable to that of $%
E_{N}^{\mathrm{out}}[0]$. The corresponding phase variation is approximately
linear in the vicinity of $\omega =0$ (as shown by the dashed linear fit in
Fig. \ref{fig:squeezingparameter}(b))%
\begin{equation}
\delta \theta \approx (\kappa /4G^{2})\delta \omega .  \label{deltatheta}
\end{equation}%
This suggests (as per Eq. (\ref{Didef})), that the optimal entanglement is
between a cavity-2 mode emitted at $\tau _{2}=0$ and a cavity-1 mode emitted
at $\tau _{1}=\kappa /4G^{2}$. The entanglement with time delay is shown by
the red-dashed curve in Fig.~3(b) of the main text. On a heuristic level, our
system first generates entangled phonon - cavity 2 photon pairs via the $%
G_{2}$ interaction in Eq. (1) in the main text; next the $G_{1}$ interaction
swaps the phonon state into a photon in cavity 1. This explains why the
optimal entanglement involves a positive delay for the cavity 1 output mode.
Finally, we note that for $\omega$ away from $0$, the squeezing phase $%
\theta $ has a nonlinear frequency dependence, and hence the optimal filter
function for larger $\sigma$ will not correspond to a simple delay as in Eq.~(\ref{deltatheta}).

\begin{figure}[tbp]
\begin{center}
\includegraphics[width=0.6\textwidth]{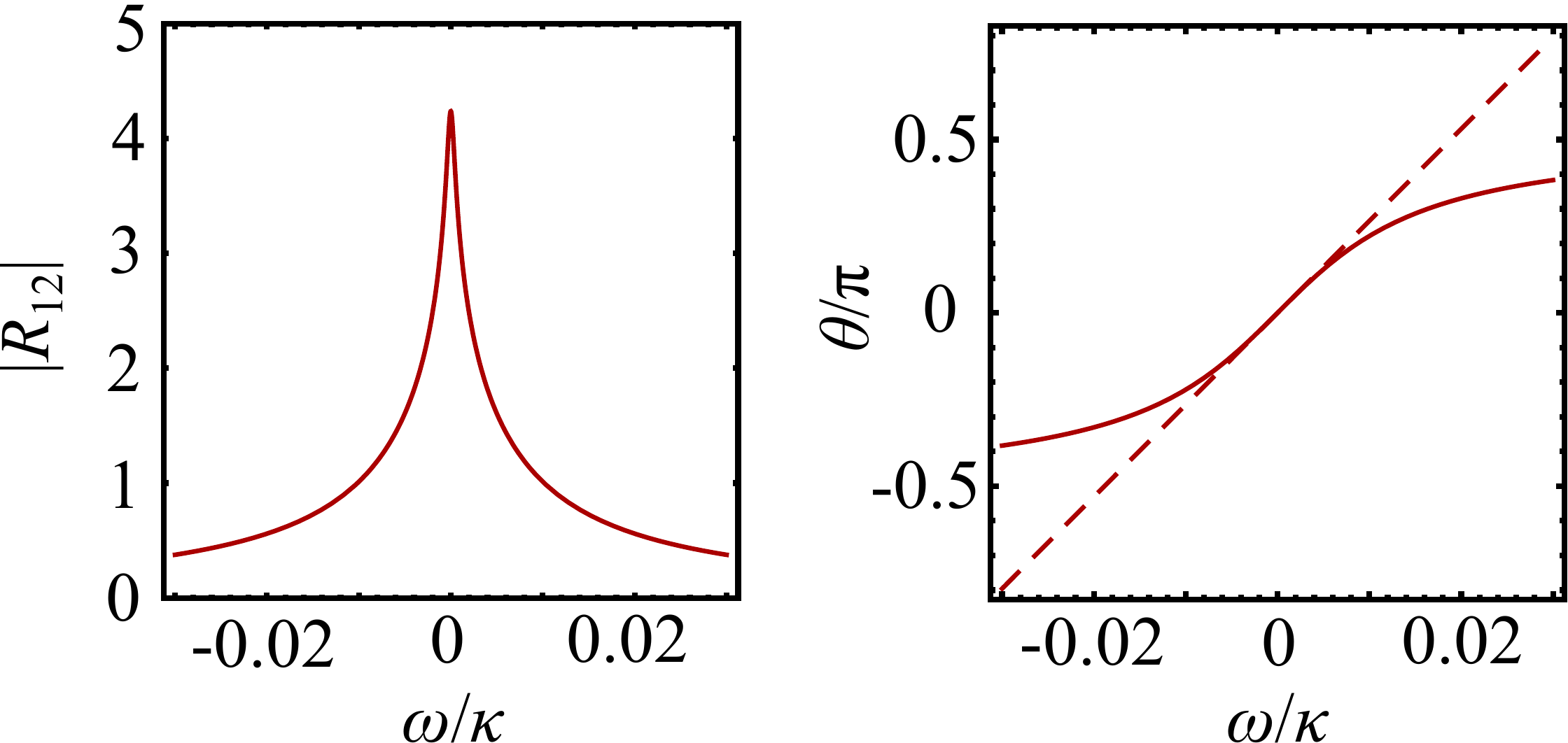}
\end{center}
\caption{The dependence of the squeezing parameter $R_{12}$ on frequency.
The left panel shows the magnitude change of $R_{12}$, while the right panel
shows the phase change of $R_{12}$. The dashed line in the right panel is
the linear fit given by Eq.~(\protect\ref{deltatheta}). The parameters used
are $G/\kappa=0.1$, $\protect\gamma/\kappa=3.3\times 10^{-5}$. }
\label{fig:squeezingparameter}
\end{figure}

\section{Phonon output through an auxiliary cavity}

In optomechanical crystals, a phonon waveguide can be used to access the
output of the mechanical system. For a general optomechanical systems where
a phonon waveguide is absent, we find it is still possible to access the
phonon output by having its damping dominated by optical damping of a third
auxiliary cavity. This auxiliary cavity will have a large damping rate, and
will be coupled to the mechanics via a linearized optomechanical coupling
(the cavity is strongly driven by a red-detuned laser):%
\begin{equation}
\hat{H}_{\mathrm{a}}=G_{\mathrm{a}}\left( \hat{b}\hat{d}_{\mathrm{a}}^{\dag
}+\hat{b}^{\dag }\hat{d}_{\mathrm{a}}\right) ,
\end{equation}%
where $\hat{d}_{\mathrm{a}}$ is the annihilation operator of the auxiliary
cavity and $G_{\mathrm{a}}$ is the corresponding coupling. The auxiliary
cavity, due to the large damping rate $\kappa _{\mathrm{a}}$, can be
described by the following steady-state relation:%
\begin{equation}
\hat{d}_{\mathrm{a}}=-i\frac{2G_{\mathrm{a}}}{\kappa _{\mathrm{a}}}\hat{b}-%
\frac{2}{\sqrt{\kappa _{\mathrm{a}}}}\hat{d}_{\mathrm{a}}^{\mathrm{in}}.
\label{da1}
\end{equation}%
Plugging this into the equation of motion for the mechanical resonator one
obtains 
\begin{equation}
\frac{d}{dt}\hat{b}=-\frac{2G_{\mathrm{a}}^{2}}{\kappa _{\mathrm{a}}}\hat{b}%
-i\left( G_{1}\hat{d}_{1}+G_{2}\hat{d}_{2}^{\dag }\right) +\frac{2iG_{%
\mathrm{a}}}{\sqrt{\kappa _{\mathrm{a}}}}\hat{d}_{\mathrm{a}}^{\mathrm{in}},
\end{equation}%
where we have neglected other damping channels of the resonator, supposing
that $G_{\mathrm{a}}^{2}/\kappa _{\mathrm{a}}$ is sufficiently large.
Comparing with the Langevin equation without auxiliary cavity $\frac{d}{dt}%
\hat{b}=-\frac{\gamma }{2}\hat{b}-i\left( G_{1}\hat{d}_{1}+G_{2}\hat{d}%
_{2}^{\dag }\right) -\sqrt{\gamma }\hat{d}_{\mathrm{a}}^{\mathrm{in}}$ we
can identify 
\begin{equation}
\hat{b}^{\mathrm{in}}=-i\hat{d}_{\mathrm{a}}^{\mathrm{in}},\text{ \ and \ }%
\gamma =4G_{\mathrm{a}}^{2}/\kappa _{\mathrm{a}}.  \label{ident_in}
\end{equation}

The input-output relation of the auxiliary cavity is $\hat{d}_{\mathrm{a}}^{%
\mathrm{out}}=\hat{d}_{\mathrm{a}}^{\mathrm{in}}+\sqrt{\kappa _{\mathrm{a}}}%
\hat{d}_{\mathrm{a}}$ which, together with Eqs. (\ref{da1}) and (\ref%
{ident_in}), gives $\hat{d}_{\mathrm{a}}^{\mathrm{out}}=-i\sqrt{\gamma }\hat{%
b}-i\hat{b}^{\mathrm{in}}$, thus 
\begin{equation}
\hat{d}_{\mathrm{a}}^{\mathrm{out}}=-i\hat{b}^{\mathrm{out}}.
\end{equation}%
This shows that a strongly damped auxiliary cavity can serve as output of
the mechanical mode.

\section{3-mode entanglement measure based on R\'{e}nyi-2 Entropy}

The tripartite entanglement can be measured by the residual Gaussian R\'{e}%
nyi-2 (GR2) entanglement~\cite{Adesso2012supp}. The R\'{e}nyi-2 entropy is given
by \ $S_{2}\left( \rho \right) =-\ln \mathrm{Tr}\left( \rho ^{2}\right) $
and allows to define an entanglement measure $\varepsilon _{2}\left( \rho
_{A:B}\right)$ for bipartite states $\rho _{AB}$. For pure states one simply
has $\varepsilon _{2}\left( \rho _{A:B}\right)=S_{2}\left( \rho _{A}\right) $%
, with $\rho _{A}$ the reduced density matrix of subsystem $A$. The
tripartite entanglement is then characterized through the residual
entanglement $\varepsilon _{2}\left( \rho _{i:j:k}\right) $, given by: 
\begin{equation}
\varepsilon _{2}\left( \rho _{i:j:k}\right) =\varepsilon _{2}\left( \rho
_{i:jk}\right) -\varepsilon _{2}\left( \rho _{i:k}\right) -\varepsilon
_{2}\left( \rho _{i:j}\right) \geq 0,  \label{Adesso1}
\end{equation}
where $i\neq j\neq k=1,2,3$ represents the 3 different modes (the 3rd mode
denotes the mechanics in our case). In Eq.~(\ref{Adesso1}), $\varepsilon
_{2}\left( \rho _{i:jk}\right) $ is the bipartite entanglement partitioning
the global system into $A=i$ and $B=jk$, while $\varepsilon _{2}\left( \rho
_{i:j}\right) $ and $\varepsilon _{2}\left( \rho _{i:k}\right) $ consider
the reduced density matrices of subsystems $ij$ and $ik$, respectively. In
general, three different values of $\varepsilon _{2}\left( \rho
_{i:j:k}\right) $ are obtained, depending on the choice of the
\textquotedblleft focus mode\textquotedblright\ $i$. There are special cases
when $\varepsilon _{2}\left( \rho _{i:j:k}\right) $ is invariant under mode
permutation \cite{Adesso2012supp}; as we explain below, that is not the case for
our system.

For pure tripartite Gaussian states, $\varepsilon _{2}\left( \rho
_{i:j:k}\right) $ can be evaluated analytically \cite{Adesso2012supp}. The first
term of Eq. (\ref{Adesso1}) is: 
\begin{equation}
\varepsilon _{2}\left( \rho _{i:jk}\right) =S_{2}\left( \rho _{i}\right)
=\ln a_{i}\text{,}
\end{equation}%
where $a_{i}$ is related to the covariance matrix $\mathbf{V}_{i}$ of
subsystem $i$: 
\begin{equation}
a_{i}=\sqrt{\det \mathbf{V}_{i}}.
\end{equation}%
Using the covariance matrix for our system, calculated from the Langevin
equation, it is possible to obtain explicit formulas for $\varepsilon
_{2}\left( \rho _{1:23}\right)$, $\varepsilon _{2}\left( \rho _{2:13}\right)$%
, $\varepsilon _{2}\left( \rho _{3:12}\right)$ in terms of the
cooperativities $C_{1,2}$. Although we omit them here, it is worth
mentioning that these three quantities are all non-zero, showing that none
of the three systems is separable.

The last two terms in Eq. (\ref{Adesso1}) are given by: 
\begin{equation*}
\varepsilon _{2}\left( \rho _{j:k}\right) =\frac{1}{2}\ln g_{i}\mathrm{\ \ \ 
}(i\neq j\neq k).
\end{equation*}%
where%
\begin{equation}
g_{i}=\left\{ 
\begin{array}{cc}
1, & a_{k}\geq \sqrt{a_{i}^{2}+a_{j}^{2}-1}, \\ 
\frac{\beta }{8a_{k}^{2}}, & \alpha _{k}<a_{k}<\sqrt{a_{i}^{2}+a_{j}^{2}-1},
\\ 
\left( \frac{a_{i}^{2}-a_{j}^{2}}{a_{k}^{2}-1}\right) ^{2} & a_{k}\leq
\alpha _{k},%
\end{array}%
\right.
\end{equation}%
with%
\begin{eqnarray*}
\alpha _{k} &=&\sqrt{\frac{2\left( a_{i}^{2}+a_{j}^{2}\right) +\left(
a_{i}^{2}-a_{j}^{2}\right) ^{2}+\left\vert a_{i}^{2}-a_{j}^{2}\right\vert 
\sqrt{\left( a_{i}^{2}-a_{j}^{2}\right) ^{2}+8\left(
a_{i}^{2}+a_{j}^{2}\right) }}{2\left( a_{i}^{2}+a_{j}^{2}\right) }}, \\
\beta
&=&2a_{1}^{2}+2a_{2}^{2}+2a_{3}^{2}+2a_{1}^{2}a_{3}^{2}+2a_{2}^{2}a_{3}^{2}+2a_{1}^{2}a_{2}^{2}-a_{1}^{4}-a_{2}^{4}-a_{3}^{4}-%
\sqrt{\delta }-1, \\
\delta &=&\left( \left( a_{1}-a_{2}-a_{3}\right) ^{2}-1\right) \left( \left(
a_{1}+a_{2}-a_{3}\right) ^{2}-1\right) \left( \left(
a_{1}-a_{2}+a_{3}\right) ^{2}-1\right) \left( \left(
a_{1}+a_{2}+a_{3}\right) ^{2}-1\right) .
\end{eqnarray*}

Since $\varepsilon _{2}\left( \rho _{j:k}\right) $ is the GR2 2-mode
entanglement measure in the subspace where the mode $i$ is eliminated, it
serves as an alternative way to quantify the 2-mode entanglement in our
system, other than the logarithmic negativity used in the main text. We
obtain all the two-mode entanglements as follows: 
\begin{eqnarray}
\varepsilon _{2}\left( \rho _{1:2}\right) &=&\ln \frac{%
(1+C_{2})^{2}+C_{1}^{2}+2C_{1}+6C_{1}C_{2}}{%
(1+C_{2})^{2}+C_{1}^{2}+2C_{1}-2C_{1}C_{2}}\approx \ln \frac{%
C_{2}^{2}+6C_{1}C_{2}+C_{1}^{2}}{\left( C_{1}-C_{2}\right) ^{2}}+\mathcal{O}%
\left( \frac{1}{C_{2}}\right) ,  \notag \\
\varepsilon _{2}\left( \rho _{3:2}\right) &=&\ln \frac{%
(1+C_{1})^{2}+C_{2}^{2}+6C_{2}+2C_{1}C_{2}}{%
(1-C_{2})^{2}+C_{1}^{2}+2C_{1}+2C_{1}C_{2}}\approx \mathcal{O}\left( \frac{1%
}{C_{2}}\right) ,  \notag \\
\varepsilon _{2}\left( \rho _{1:3}\right) &=&0,  \label{E2_Reny}
\end{eqnarray}%
where the last approximation is in the large $C$ limit. These results show
that the entanglement between cavity 1 and the mechanics is always zero,
although the mechanics is entangled with the composite system of the two
cavities ($\varepsilon _{2}\left( \rho _{3:12}\right)\neq 0$). In
particular, there is entanglement between the mechanics and cavity 2,
although it is much smaller than the entanglement between cavity 1 and 2.
These results are all in agreement with those based on the logarithmic
negativity.

We now turn to the evaluation of the tripartite entanglement $\varepsilon
_{2}\left( \rho _{i:j:k}\right) $, which is permutationally invariant only
when all the $\varepsilon _{2}\left( \rho _{i:jk}\right) $ and $\varepsilon
_{2}\left( \rho _{i:j}\right) $ are non-zero \cite{Adesso2012supp}. The fact
that $\varepsilon _{2}\left( \rho _{1:3}\right) =0$ implies that $%
\varepsilon _{2}\left( \rho _{i:j:k}\right) $ depends on the focus mode,
thus we consider all three choices $i=1,2,3$. Using the results for $%
\varepsilon _{2}\left( \rho _{i:jk}\right) $, $\varepsilon _{2}\left( \rho
_{i:j}\right)$ discussed above, Eq.~(\ref{Adesso1}) gives: 
\begin{eqnarray}
\varepsilon _{2}\left( \rho _{1:2:3}\right) &=&\ln \left( \frac{%
C_{1}^{2}+\left( 1+C_{2}\right) ^{2}+2C_{1}-2C_{1}C_{2}}{\left(
1+C_{1}-C_{2}\right) ^{2}}\right) -2\tanh ^{-1}\left( \frac{2C_{2}}{\left(
1+C_{1}\right) ^{2}+6C_{1}C_{2}+C_{2}^{2}}\right) ,  \notag \\
\varepsilon _{2}\left( \rho _{2:1:3}\right) &=&\ln \left( \frac{%
C_{2}^{2}+\left( 1+C_{1}\right) ^{2}+6C_{2}\left( 1+C_{1}\right) }{%
C_{2}^{2}+\left( 1+C_{1}\right) ^{2}+2C_{2}\left( 3+C_{1}\right) }\right) 
\notag \\
&&+\ln \left( \frac{\left( C_{1}^{2}+\left( 1+C_{2}\right) ^{2}-2C_{1}\left(
C_{2}-1\right) \right) \left( C_{1}^{2}+\left( 1-C_{2}\right)
^{2}+2C_{1}\left( 1+C_{2}\right) \right) }{\left( 1+C_{1}-C_{2}\right)
^{2}\left( C_{1}^{2}+\left( 1+C_{2}\right) ^{2}+2C_{1}\left( 1+3C_{2}\right)
\right) }\right) ,  \notag \\
\varepsilon _{2}\left( \rho _{3:1:2}\right) &=&\ln \left( \frac{%
C_{2}^{2}+\left( 1+C_{1}\right) ^{2}+6C_{2}-2C_{1}C_{2}}{\left(
1+C_{1}-C_{2}\right) ^{2}}\right) -2\tanh ^{-1}\left( \frac{4C_{2}}{\left(
1+C_{1}-C_{2}\right) ^{2}}\right) .
\end{eqnarray}%
As announced, the three \textquotedblleft residual GR2
entanglements\textquotedblright\ are unequal. Nevertheless, they are all
larger than zero, which confirms the presence of genuine tripartite
entanglement. Furthermore, as shown in Fig.~\ref{fig:3mode}, the $%
\varepsilon _{2}\left( \rho _{i:j:k}\right) $ all diverge at the instability
point $\gamma _{\mathrm{tot}}=0$. This divergence is similar to the case of
a parametric amplifier and is related to the divergences of $\bar{n}_{2}$
and $R_{2\mathrm{m}}$ as discussed in the main text. 
\begin{figure}[tbp]
\begin{center}
\includegraphics[width=0.4\textwidth]{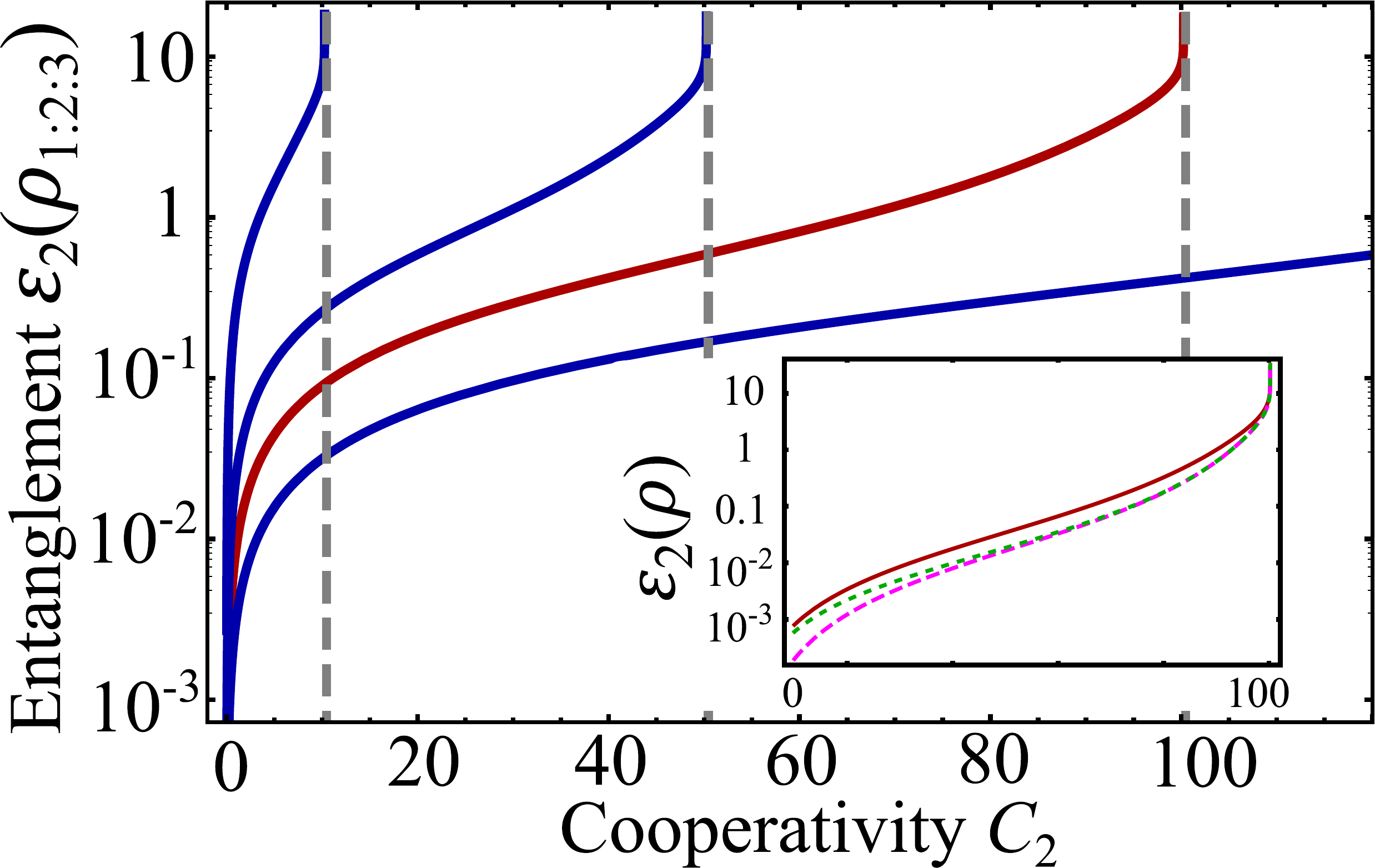}
\end{center}
\caption{3-mode entanglement, evaluated by the residual GR2 entanglement
with cavity-1 as the focus mode $\protect\epsilon _{2}(\protect\rho %
_{1:2:3}) $, versus the cooperativity of cavity 2. Here $C_{1}$ is fixed to $%
10,\ 50,\ 100,\ 200$ (from top to bottom). The gray dashed lines indicate
the onset of system instability. The 3-mode entanglement diverges at $%
\protect\gamma _{\mathrm{tot}}=0$ or $C_{2}=C_{1}+1$. The $C_{1}=100$ (red)
curve is re-plotted in the inset, together with the two residual GR2
entanglement where the focus mode is cavity-2 ($\protect\epsilon _{2}(%
\protect\rho _{2:1:3})$) or mechanics ($\protect\epsilon _{2}(\protect\rho %
_{3:2:1})$).}
\label{fig:3mode}
\end{figure}

\section{The tripartite entangled output state}

We consider the 3-mode output state $\left\vert \Psi \right\rangle $
(including both cavities and the mechanical resonator) at $\omega =0$ and $%
\sigma =0$. Since $\bar{n}_{1}=0$, see Eq. (7) of the main text, cavity $1$
must be in the ground state if we consider the ``unsqueezed" state $%
\left\vert \Psi _{s}\right\rangle \equiv \tilde{S}_{12}\left( -R_{12}\right)
\left\vert \Psi \right\rangle $. However, $\left\vert \Psi _{s}\right\rangle 
$ has residual entanglement between cavity $2$ and the mechanical mode. This
is clear from the finite value of $\bar{n}_{2}$, which diverges at the
instability condition $\gamma _{\mathrm{tot}}=0$ (the finite population $%
\bar{n}_{2}$ is a consequence of tracing out the mechanical mode). One can
compute explicitly the correlations of the state $\left\vert \Psi
_{s}\right\rangle $ and identify it as a squeezed vacuum of cavity 2 and the
mechanical mode, with squeezing parameter $R_{2\mathrm{m}}=\sinh ^{-1}\bar{n}%
_{2}$. Hence it can be concluded that 
\begin{equation}
\left\vert \Psi \right\rangle =\hat{S}_{12}\left( R_{12}\right) \hat{S}_{2%
\mathrm{m}}\left( R_{2\mathrm{m}}\right) \left\vert 000\right\rangle ,
\label{3modessupp}
\end{equation}%
with 
\begin{equation}
\hat{S}_{2\mathrm{m}}\left( R_{2\mathrm{m}}\right) \equiv \exp \left[ iR_{2%
\mathrm{m}}\hat{D}_{\mathrm{m}}^{\mathrm{out}}\left[ 0\right] \hat{D}_{2}^{%
\mathrm{out}}\left[ 0\right] -h.c.\right] .  \label{defr1}
\end{equation}

Another interesting observation is that, although the tripartite steady
state Eq. (\ref{3modessupp}) is generated by interactions between intracavity
modes and the input-output relations, the output state described by Eq.~(\ref%
{3modessupp}) could also be achieved by a unitary evolution generated by the
following interaction between the output modes:
\begin{equation}
\tilde{H}=R_{12}\sin R_{2\mathrm{m}}\hat{D}_{\mathrm{m}}^{\mathrm{out}}\left[
0\right] \hat{D}_{2}^{\mathrm{out}}\left[ 0\right] -R_{2\mathrm{m}}\left( 
\hat{D}_{\mathrm{m}}^{\mathrm{out}}\left[ 0\right] \right) ^{\dag }\hat{D}%
_{1}^{\mathrm{out}}\left[ 0\right] +iR_{12}\cos R_{2\mathrm{m}}\hat{D}_{1}^{%
\mathrm{out}}\left[ 0\right] \hat{D}_{2}^{\mathrm{out}}\left[ 0\right] +h.c.
\end{equation}%
This result can be obtained by Wei-Norman decomposition of $\exp \left( -i%
\tilde{H}\right) $ in terms of the closed algebra $\left\{ \left( \hat{D}_{%
\mathrm{m}}^{\mathrm{out}}\left[ 0\right] \right) ^{\dag }\hat{D}_{1}^{%
\mathrm{out}}\left[ 0\right] +h.c.,\hat{D}_{\mathrm{m}}^{\mathrm{out}}\left[
0\right] \hat{D}_{1}^{\mathrm{out}}\left[ 0\right] +h.c.,\hat{D}_{1}^{%
\mathrm{out}}\left[ 0\right] \hat{D}_{2}^{\mathrm{out}}\left[ 0\right]
-h.c.\right\} $.

Using this effective Hamiltonian, we can derive the expansion of the output
state into the Fock state basis:%
\begin{eqnarray}
\left\vert \Psi \right\rangle &=&\exp \left( -i\tilde{H}\right) \left\vert
000\right\rangle \\
&=&\frac{1}{\sqrt{1+\left\langle N_{2}\right\rangle }}\sum_{pq}C_{p+q}^{p}%
\left( \frac{\left\langle N_{\mathrm{m}}\right\rangle }{1+\left\langle
N_{2}\right\rangle }\right) ^{\frac{q}{2}}\left( \frac{\left\langle
N_{1}\right\rangle }{1+\left\langle N_{2}\right\rangle }\right) ^{\frac{p}{2}%
}\left\vert p,p+q,q\right\rangle ,  \label{fock_3mode}
\end{eqnarray}%
where 
\begin{eqnarray}
\left\langle \hat{N}_{1}\right\rangle &=&\frac{4C_{1}C_{2}}{\left(
1+C_{1}-C_{2}\right) ^{2}},  \notag \\
\left\langle \hat{N}_{2}\right\rangle &=&\frac{4C_{2}\left( C_{1}+1\right) }{%
\left( 1+C_{1}-C_{2}\right) ^{2}},  \notag \\
\left\langle \hat{N}_{m}\right\rangle &=&\frac{4C_{2}}{\left(
1+C_{1}-C_{2}\right) ^{2}},
\end{eqnarray}%
with $p$, $q$ integers and $\hat{N}_{i}\equiv \left( \hat{D}_{i}^{\mathrm{out%
}}\left[ 0\right] \right) ^{\dag }\hat{D}_{i}^{\mathrm{out}}\left[ 0\right] $
($i=1,2,\mathrm{m}$). The genuine tripartite entanglement is evident from
the perfect correlations between photon and phonon numbers in the output
state of Eq. (\ref{fock_3mode}).

\end{widetext}


\begin{thebibliography}{37}%
\makeatletter
\providecommand \@ifxundefined [1]{%
 \@ifx{#1\undefined}
}%
\providecommand \@ifnum [1]{%
 \ifnum #1\expandafter \@firstoftwo
 \else \expandafter \@secondoftwo
 \fi
}%
\providecommand \@ifx [1]{%
 \ifx #1\expandafter \@firstoftwo
 \else \expandafter \@secondoftwo
 \fi
}%
\providecommand \natexlab [1]{#1}%
\providecommand \enquote  [1]{``#1''}%
\providecommand \bibnamefont  [1]{#1}%
\providecommand \bibfnamefont [1]{#1}%
\providecommand \citenamefont [1]{#1}%
\providecommand \href@noop [0]{\@secondoftwo}%
\providecommand \href [0]{\begingroup \@sanitize@url \@href}%
\providecommand \@href[1]{\@@startlink{#1}\@@href}%
\providecommand \@@href[1]{\endgroup#1\@@endlink}%
\providecommand \@sanitize@url [0]{\catcode `\\12\catcode `\$12\catcode
  `\&12\catcode `\#12\catcode `\^12\catcode `\_12\catcode `\%12\relax}%
\providecommand \@@startlink[1]{}%
\providecommand \@@endlink[0]{}%
\providecommand \url  [0]{\begingroup\@sanitize@url \@url }%
\providecommand \@url [1]{\endgroup\@href {#1}{\urlprefix }}%
\providecommand \urlprefix  [0]{URL }%
\providecommand \Eprint [0]{\href }%
\providecommand \doibase [0]{http://dx.doi.org/}%
\providecommand \selectlanguage [0]{\@gobble}%
\providecommand \bibinfo  [0]{\@secondoftwo}%
\providecommand \bibfield  [0]{\@secondoftwo}%
\providecommand \translation [1]{[#1]}%
\providecommand \BibitemOpen [0]{}%
\providecommand \bibitemStop [0]{}%
\providecommand \bibitemNoStop [0]{.\EOS\space}%
\providecommand \EOS [0]{\spacefactor3000\relax}%
\providecommand \BibitemShut  [1]{\csname bibitem#1\endcsname}%
\let\auto@bib@innerbib\@empty
\bibitem [{\citenamefont {Aoki}\ \emph {et~al.}(2003)\citenamefont {Aoki},
  \citenamefont {Takei}, \citenamefont {Yonezawa}, \citenamefont {Wakui},
  \citenamefont {Hiraoka}, \citenamefont {Furusawa},\ and\ \citenamefont {van
  Loock}}]{Aoki2003}%
  \BibitemOpen
  \bibfield  {author} {\bibinfo {author} {\bibfnamefont {T.}~\bibnamefont
  {Aoki}}, \bibinfo {author} {\bibfnamefont {N.}~\bibnamefont {Takei}},
  \bibinfo {author} {\bibfnamefont {H.}~\bibnamefont {Yonezawa}}, \bibinfo
  {author} {\bibfnamefont {K.}~\bibnamefont {Wakui}}, \bibinfo {author}
  {\bibfnamefont {T.}~\bibnamefont {Hiraoka}}, \bibinfo {author} {\bibfnamefont
  {A.}~\bibnamefont {Furusawa}}, \ and\ \bibinfo {author} {\bibfnamefont
  {P.}~\bibnamefont {van Loock}},\ }\href {\doibase
  10.1103/PhysRevLett.91.080404} {\bibfield  {journal} {\bibinfo  {journal}
  {Phys. Rev. Lett.}\ }\textbf {\bibinfo {volume} {91}},\ \bibinfo {pages}
  {080404} (\bibinfo {year} {2003})}\BibitemShut {NoStop}%
\bibitem [{\citenamefont {Pan}\ \emph {et~al.}(2012)\citenamefont {Pan},
  \citenamefont {Chen}, \citenamefont {Lu}, \citenamefont {Weinfurter},
  \citenamefont {Zeilinger},\ and\ \citenamefont {\ifmmode~\dot{Z}\else
  \.{Z}\fi{}ukowski}}]{Pan2012}%
  \BibitemOpen
  \bibfield  {author} {\bibinfo {author} {\bibfnamefont {J.-W.}\ \bibnamefont
  {Pan}}, \bibinfo {author} {\bibfnamefont {Z.-B.}\ \bibnamefont {Chen}},
  \bibinfo {author} {\bibfnamefont {C.-Y.}\ \bibnamefont {Lu}}, \bibinfo
  {author} {\bibfnamefont {H.}~\bibnamefont {Weinfurter}}, \bibinfo {author}
  {\bibfnamefont {A.}~\bibnamefont {Zeilinger}}, \ and\ \bibinfo {author}
  {\bibfnamefont {M.}~\bibnamefont {\ifmmode~\dot{Z}\else \.{Z}\fi{}ukowski}},\
  }\href {\doibase 10.1103/RevModPhys.84.777} {\bibfield  {journal} {\bibinfo
  {journal} {Rev. Mod. Phys.}\ }\textbf {\bibinfo {volume} {84}},\ \bibinfo
  {pages} {777} (\bibinfo {year} {2012})}\BibitemShut {NoStop}%
\bibitem [{\citenamefont {Raimond}\ \emph {et~al.}(2001)\citenamefont
  {Raimond}, \citenamefont {Brune},\ and\ \citenamefont
  {Haroche}}]{Raimond2001}%
  \BibitemOpen
  \bibfield  {author} {\bibinfo {author} {\bibfnamefont {J.}~\bibnamefont
  {Raimond}}, \bibinfo {author} {\bibfnamefont {M.}~\bibnamefont {Brune}}, \
  and\ \bibinfo {author} {\bibfnamefont {S.}~\bibnamefont {Haroche}},\
  }\href@noop {} {\bibfield  {journal} {\bibinfo  {journal} {Rev. Mod. Phys.}\
  }\textbf {\bibinfo {volume} {73}},\ \bibinfo {pages} {565} (\bibinfo {year}
  {2001})}\BibitemShut {NoStop}%
\bibitem [{\citenamefont {Neeley}\ \emph {et~al.}(2010)\citenamefont {Neeley},
  \citenamefont {Bialczak}, \citenamefont {Lenander}, \citenamefont {Lucero},
  \citenamefont {Mariantoni}, \citenamefont {Sank}, \citenamefont {Wang},
  \citenamefont {Weides}, \citenamefont {Wenner}, \citenamefont {Yin},
  \citenamefont {Yamamoto}, \citenamefont {Cleland},\ and\ \citenamefont
  {Martinis}}]{Neeley2010}%
  \BibitemOpen
  \bibfield  {author} {\bibinfo {author} {\bibfnamefont {M.}~\bibnamefont
  {Neeley}}, \bibinfo {author} {\bibfnamefont {R.~C.}\ \bibnamefont
  {Bialczak}}, \bibinfo {author} {\bibfnamefont {M.}~\bibnamefont {Lenander}},
  \bibinfo {author} {\bibfnamefont {E.}~\bibnamefont {Lucero}}, \bibinfo
  {author} {\bibfnamefont {M.}~\bibnamefont {Mariantoni}}, \bibinfo {author}
  {\bibfnamefont {D.}~\bibnamefont {Sank}}, \bibinfo {author} {\bibfnamefont
  {H.}~\bibnamefont {Wang}}, \bibinfo {author} {\bibfnamefont {M.}~\bibnamefont
  {Weides}}, \bibinfo {author} {\bibfnamefont {J.}~\bibnamefont {Wenner}},
  \bibinfo {author} {\bibfnamefont {Y.}~\bibnamefont {Yin}}, \bibinfo {author}
  {\bibfnamefont {T.}~\bibnamefont {Yamamoto}}, \bibinfo {author}
  {\bibfnamefont {A.~N.}\ \bibnamefont {Cleland}}, \ and\ \bibinfo {author}
  {\bibfnamefont {J.~M.}\ \bibnamefont {Martinis}},\ }\href@noop {} {\bibfield
  {journal} {\bibinfo  {journal} {Nature}\ }\textbf {\bibinfo {volume} {467}},\
  \bibinfo {pages} {570} (\bibinfo {year} {2010})}\BibitemShut {NoStop}%
\bibitem [{\citenamefont {DiCarlo}\ \emph {et~al.}(2010)\citenamefont
  {DiCarlo}, \citenamefont {Reed}, \citenamefont {Sun}, \citenamefont
  {Johnson}, \citenamefont {Chow}, \citenamefont {Gambetta}, \citenamefont
  {Frunzio}, \citenamefont {Girvin}, \citenamefont {Devoret},\ and\
  \citenamefont {Schoelkopf}}]{Dicarlo2010}%
  \BibitemOpen
  \bibfield  {author} {\bibinfo {author} {\bibfnamefont {L.}~\bibnamefont
  {DiCarlo}}, \bibinfo {author} {\bibfnamefont {M.}~\bibnamefont {Reed}},
  \bibinfo {author} {\bibfnamefont {L.}~\bibnamefont {Sun}}, \bibinfo {author}
  {\bibfnamefont {B.~L.}\ \bibnamefont {Johnson}}, \bibinfo {author}
  {\bibfnamefont {J.~M.}\ \bibnamefont {Chow}}, \bibinfo {author}
  {\bibfnamefont {J.~M.}\ \bibnamefont {Gambetta}}, \bibinfo {author}
  {\bibfnamefont {L.}~\bibnamefont {Frunzio}}, \bibinfo {author} {\bibfnamefont
  {S.~M.}\ \bibnamefont {Girvin}}, \bibinfo {author} {\bibfnamefont {M.~H.}\
  \bibnamefont {Devoret}}, \ and\ \bibinfo {author} {\bibfnamefont {R.~J.}\
  \bibnamefont {Schoelkopf}},\ }\href@noop {} {\bibfield  {journal} {\bibinfo
  {journal} {Nature}\ }\textbf {\bibinfo {volume} {467}},\ \bibinfo {pages}
  {574} (\bibinfo {year} {2010})}\BibitemShut {NoStop}%
\bibitem [{\citenamefont {Flurin}\ \emph {et~al.}(2012)\citenamefont {Flurin},
  \citenamefont {Roch}, \citenamefont {Mallet}, \citenamefont {Devoret},\ and\
  \citenamefont {Huard}}]{Flurin2012}%
  \BibitemOpen
  \bibfield  {author} {\bibinfo {author} {\bibfnamefont {E.}~\bibnamefont
  {Flurin}}, \bibinfo {author} {\bibfnamefont {N.}~\bibnamefont {Roch}},
  \bibinfo {author} {\bibfnamefont {F.}~\bibnamefont {Mallet}}, \bibinfo
  {author} {\bibfnamefont {M.~H.}\ \bibnamefont {Devoret}}, \ and\ \bibinfo
  {author} {\bibfnamefont {B.}~\bibnamefont {Huard}},\ }\href {\doibase
  10.1103/PhysRevLett.109.183901} {\bibfield  {journal} {\bibinfo  {journal}
  {Phys. Rev. Lett.}\ }\textbf {\bibinfo {volume} {109}},\ \bibinfo {pages}
  {183901} (\bibinfo {year} {2012})}\BibitemShut {NoStop}%
\bibitem [{\citenamefont {Bernien}\ \emph {et~al.}(2013)\citenamefont
  {Bernien}, \citenamefont {Hensen}, \citenamefont {Pfaff}, \citenamefont
  {Koolstra}, \citenamefont {Blok}, \citenamefont {Robledo}, \citenamefont
  {Taminiau}, \citenamefont {Markham}, \citenamefont {Twitchen}, \citenamefont
  {Childress},\ and\ \citenamefont {Hanson}}]{Bernien2013}%
  \BibitemOpen
  \bibfield  {author} {\bibinfo {author} {\bibfnamefont {H.}~\bibnamefont
  {Bernien}}, \bibinfo {author} {\bibfnamefont {B.}~\bibnamefont {Hensen}},
  \bibinfo {author} {\bibfnamefont {W.}~\bibnamefont {Pfaff}}, \bibinfo
  {author} {\bibfnamefont {G.}~\bibnamefont {Koolstra}}, \bibinfo {author}
  {\bibfnamefont {M.~S.}\ \bibnamefont {Blok}}, \bibinfo {author}
  {\bibfnamefont {L.}~\bibnamefont {Robledo}}, \bibinfo {author} {\bibfnamefont
  {T.~H.}\ \bibnamefont {Taminiau}}, \bibinfo {author} {\bibfnamefont
  {M.}~\bibnamefont {Markham}}, \bibinfo {author} {\bibfnamefont {D.~J.}\
  \bibnamefont {Twitchen}}, \bibinfo {author} {\bibfnamefont {L.}~\bibnamefont
  {Childress}}, \ and\ \bibinfo {author} {\bibfnamefont {R.}~\bibnamefont
  {Hanson}},\ }\href@noop {} {\bibfield  {journal} {\bibinfo  {journal}
  {Nature}\ }\textbf {\bibinfo {volume} {497}},\ \bibinfo {pages} {86}
  (\bibinfo {year} {2013})}\BibitemShut {NoStop}%
\bibitem [{\citenamefont {Aspelmeyer}\ \emph {et~al.}(2013)\citenamefont
  {Aspelmeyer}, \citenamefont {Kippenberg},\ and\ \citenamefont
  {Marquardt}}]{FlorianRMP}%
  \BibitemOpen
  \bibfield  {author} {\bibinfo {author} {\bibfnamefont {M.}~\bibnamefont
  {Aspelmeyer}}, \bibinfo {author} {\bibfnamefont {T.~J.}\ \bibnamefont
  {Kippenberg}}, \ and\ \bibinfo {author} {\bibfnamefont {F.}~\bibnamefont
  {Marquardt}},\ }\href@noop {} {\bibfield  {journal} {\bibinfo  {journal}
  {arXiv:1303.0733}\ } (\bibinfo {year} {2013})}\BibitemShut {NoStop}%
\bibitem [{\citenamefont {Paternostro}\ \emph {et~al.}(2007)\citenamefont
  {Paternostro}, \citenamefont {Vitali}, \citenamefont {Gigan}, \citenamefont
  {Kim}, \citenamefont {Brukner}, \citenamefont {Eisert},\ and\ \citenamefont
  {Aspelmeyer}}]{Paternostro2007}%
  \BibitemOpen
  \bibfield  {author} {\bibinfo {author} {\bibfnamefont {M.}~\bibnamefont
  {Paternostro}}, \bibinfo {author} {\bibfnamefont {D.}~\bibnamefont {Vitali}},
  \bibinfo {author} {\bibfnamefont {S.}~\bibnamefont {Gigan}}, \bibinfo
  {author} {\bibfnamefont {M.}~\bibnamefont {Kim}}, \bibinfo {author}
  {\bibfnamefont {C.}~\bibnamefont {Brukner}}, \bibinfo {author} {\bibfnamefont
  {J.}~\bibnamefont {Eisert}}, \ and\ \bibinfo {author} {\bibfnamefont
  {M.}~\bibnamefont {Aspelmeyer}},\ }\href@noop {} {\bibfield  {journal}
  {\bibinfo  {journal} {Phys. Rev. Lett.}\ }\textbf {\bibinfo {volume} {99}},\
  \bibinfo {pages} {250401} (\bibinfo {year} {2007})}\BibitemShut {NoStop}%
\bibitem [{\citenamefont {Genes}\ \emph {et~al.}(2008)\citenamefont {Genes},
  \citenamefont {Mari}, \citenamefont {Tombesi},\ and\ \citenamefont
  {Vitali}}]{Genes2008}%
  \BibitemOpen
  \bibfield  {author} {\bibinfo {author} {\bibfnamefont {C.}~\bibnamefont
  {Genes}}, \bibinfo {author} {\bibfnamefont {A.}~\bibnamefont {Mari}},
  \bibinfo {author} {\bibfnamefont {P.}~\bibnamefont {Tombesi}}, \ and\
  \bibinfo {author} {\bibfnamefont {D.}~\bibnamefont {Vitali}},\ }\href@noop {}
  {\bibfield  {journal} {\bibinfo  {journal} {Phys. Rev. A}\ }\textbf {\bibinfo
  {volume} {78}},\ \bibinfo {pages} {032316} (\bibinfo {year}
  {2008})}\BibitemShut {NoStop}%
\bibitem [{\citenamefont {Wipf}\ \emph {et~al.}(2008)\citenamefont {Wipf},
  \citenamefont {Corbitt}, \citenamefont {Chen},\ and\ \citenamefont
  {Mavalvala}}]{Wipf2008}%
  \BibitemOpen
  \bibfield  {author} {\bibinfo {author} {\bibfnamefont {C.}~\bibnamefont
  {Wipf}}, \bibinfo {author} {\bibfnamefont {T.}~\bibnamefont {Corbitt}},
  \bibinfo {author} {\bibfnamefont {Y.}~\bibnamefont {Chen}}, \ and\ \bibinfo
  {author} {\bibfnamefont {N.}~\bibnamefont {Mavalvala}},\ }\href@noop {}
  {\bibfield  {journal} {\bibinfo  {journal} {New J. Phys.}\ }\textbf {\bibinfo
  {volume} {10}} (\bibinfo {year} {2008})}\BibitemShut {NoStop}%
\bibitem [{\citenamefont {Hofer}\ \emph {et~al.}(2011)\citenamefont {Hofer},
  \citenamefont {Wieczorek}, \citenamefont {Aspelmeyer},\ and\ \citenamefont
  {Hammerer}}]{Hofer2011}%
  \BibitemOpen
  \bibfield  {author} {\bibinfo {author} {\bibfnamefont {S.~G.}\ \bibnamefont
  {Hofer}}, \bibinfo {author} {\bibfnamefont {W.}~\bibnamefont {Wieczorek}},
  \bibinfo {author} {\bibfnamefont {M.}~\bibnamefont {Aspelmeyer}}, \ and\
  \bibinfo {author} {\bibfnamefont {K.}~\bibnamefont {Hammerer}},\ }\href
  {\doibase 10.1103/PhysRevA.84.052327} {\bibfield  {journal} {\bibinfo
  {journal} {Phys. Rev. A}\ }\textbf {\bibinfo {volume} {84}},\ \bibinfo
  {pages} {052327} (\bibinfo {year} {2011})}\BibitemShut {NoStop}%
\bibitem [{\citenamefont {Barzanjeh}\ \emph {et~al.}(2012)\citenamefont
  {Barzanjeh}, \citenamefont {Abdi}, \citenamefont {Milburn}, \citenamefont
  {Tombesi},\ and\ \citenamefont {Vitali}}]{Barzanjeh2012}%
  \BibitemOpen
  \bibfield  {author} {\bibinfo {author} {\bibfnamefont {S.}~\bibnamefont
  {Barzanjeh}}, \bibinfo {author} {\bibfnamefont {M.}~\bibnamefont {Abdi}},
  \bibinfo {author} {\bibfnamefont {G.}~\bibnamefont {Milburn}}, \bibinfo
  {author} {\bibfnamefont {P.}~\bibnamefont {Tombesi}}, \ and\ \bibinfo
  {author} {\bibfnamefont {D.}~\bibnamefont {Vitali}},\ }\href@noop {}
  {\bibfield  {journal} {\bibinfo  {journal} {Phys. Rev. Lett.}\ }\textbf
  {\bibinfo {volume} {109}},\ \bibinfo {pages} {130503} (\bibinfo {year}
  {2012})}\BibitemShut {NoStop}%
\bibitem [{\citenamefont {Tian}(2013)}]{Tian2013}%
  \BibitemOpen
  \bibfield  {author} {\bibinfo {author} {\bibfnamefont {L.}~\bibnamefont
  {Tian}},\ }\href {\doibase 10.1103/PhysRevLett.110.233602} {\bibfield
  {journal} {\bibinfo  {journal} {Phys. Rev. Lett.}\ }\textbf {\bibinfo
  {volume} {110}},\ \bibinfo {pages} {233602} (\bibinfo {year}
  {2013})}\BibitemShut {NoStop}%
\bibitem [{\citenamefont {Kuzyk}\ \emph {et~al.}(2013)\citenamefont {Kuzyk},
  \citenamefont {van Enk},\ and\ \citenamefont {Wang}}]{Kuzyk2013}%
  \BibitemOpen
  \bibfield  {author} {\bibinfo {author} {\bibfnamefont {M.~C.}\ \bibnamefont
  {Kuzyk}}, \bibinfo {author} {\bibfnamefont {S.~J.}\ \bibnamefont {van Enk}},
  \ and\ \bibinfo {author} {\bibfnamefont {H.}~\bibnamefont {Wang}},\
  }\href@noop {} {\bibfield  {journal} {\bibinfo  {journal} {Phys. Rev. A}\
  }\textbf {\bibinfo {volume} {88}},\ \bibinfo {pages} {062341} (\bibinfo
  {year} {2013})}\BibitemShut {NoStop}%
\bibitem [{\citenamefont {Palomaki}\ \emph {et~al.}(2013)\citenamefont
  {Palomaki}, \citenamefont {Teufel}, \citenamefont {Simmonds},\ and\
  \citenamefont {Lehnert}}]{Palomaki2013}%
  \BibitemOpen
  \bibfield  {author} {\bibinfo {author} {\bibfnamefont {T.~A.}\ \bibnamefont
  {Palomaki}}, \bibinfo {author} {\bibfnamefont {J.~D.}\ \bibnamefont
  {Teufel}}, \bibinfo {author} {\bibfnamefont {R.~W.}\ \bibnamefont
  {Simmonds}}, \ and\ \bibinfo {author} {\bibfnamefont {K.~W.}\ \bibnamefont
  {Lehnert}},\ }\href {\doibase 10.1126/science.1244563} {\bibfield  {journal}
  {\bibinfo  {journal} {Science}\ }\textbf {\bibinfo {volume} {342}},\ \bibinfo
  {pages} {710} (\bibinfo {year} {2013})}\BibitemShut {NoStop}%
\bibitem [{\citenamefont {Dong}\ \emph {et~al.}(2012)\citenamefont {Dong},
  \citenamefont {Fiore}, \citenamefont {Kuzyk},\ and\ \citenamefont
  {Wang}}]{Dong2012}%
  \BibitemOpen
  \bibfield  {author} {\bibinfo {author} {\bibfnamefont {C.}~\bibnamefont
  {Dong}}, \bibinfo {author} {\bibfnamefont {V.}~\bibnamefont {Fiore}},
  \bibinfo {author} {\bibfnamefont {M.~C.}\ \bibnamefont {Kuzyk}}, \ and\
  \bibinfo {author} {\bibfnamefont {H.}~\bibnamefont {Wang}},\ }\href@noop {}
  {\bibfield  {journal} {\bibinfo  {journal} {Science}\ }\textbf {\bibinfo
  {volume} {338}},\ \bibinfo {pages} {1609} (\bibinfo {year}
  {2012})}\BibitemShut {NoStop}%
\bibitem [{\citenamefont {Hill}\ \emph {et~al.}(2012)\citenamefont {Hill},
  \citenamefont {Safavi-Naeini}, \citenamefont {Chan},\ and\ \citenamefont
  {Painter}}]{Hill2012}%
  \BibitemOpen
  \bibfield  {author} {\bibinfo {author} {\bibfnamefont {J.~T.}\ \bibnamefont
  {Hill}}, \bibinfo {author} {\bibfnamefont {A.~H.}\ \bibnamefont
  {Safavi-Naeini}}, \bibinfo {author} {\bibfnamefont {J.}~\bibnamefont {Chan}},
  \ and\ \bibinfo {author} {\bibfnamefont {O.}~\bibnamefont {Painter}},\
  }\href@noop {} {\bibfield  {journal} {\bibinfo  {journal} {Nat. Commun.}\
  }\textbf {\bibinfo {volume} {3}},\ \bibinfo {pages} {1196} (\bibinfo {year}
  {2012})}\BibitemShut {NoStop}%
\bibitem [{\citenamefont {Andrews}\ \emph {et~al.}(2014)\citenamefont
  {Andrews}, \citenamefont {Peterson}, \citenamefont {Purdy}, \citenamefont
  {Cicak}, \citenamefont {Simmonds}, \citenamefont {Regal},\ and\ \citenamefont
  {Lehnert}}]{Andrews2014}%
  \BibitemOpen
  \bibfield  {author} {\bibinfo {author} {\bibfnamefont {R.}~\bibnamefont
  {Andrews}}, \bibinfo {author} {\bibfnamefont {R.~W.}\ \bibnamefont
  {Peterson}}, \bibinfo {author} {\bibfnamefont {T.~P.}\ \bibnamefont {Purdy}},
  \bibinfo {author} {\bibfnamefont {K.}~\bibnamefont {Cicak}}, \bibinfo
  {author} {\bibfnamefont {R.~W.}\ \bibnamefont {Simmonds}}, \bibinfo {author}
  {\bibfnamefont {C.~A.}\ \bibnamefont {Regal}}, \ and\ \bibinfo {author}
  {\bibfnamefont {K.~W.}\ \bibnamefont {Lehnert}},\ }\href@noop {} {\bibfield
  {journal} {\bibinfo  {journal} {Nat. Phys.}\ }\textbf {\bibinfo {volume}
  {10}},\ \bibinfo {pages} {321} (\bibinfo {year} {2014})}\BibitemShut
  {NoStop}%
\bibitem [{\citenamefont {Safavi-Naeini}\ and\ \citenamefont
  {Painter}(2011)}]{Safavi2011}%
  \BibitemOpen
  \bibfield  {author} {\bibinfo {author} {\bibfnamefont {A.~H.}\ \bibnamefont
  {Safavi-Naeini}}\ and\ \bibinfo {author} {\bibfnamefont {O.}~\bibnamefont
  {Painter}},\ }\href@noop {} {\bibfield  {journal} {\bibinfo  {journal} {New
  J. Phys.}\ }\textbf {\bibinfo {volume} {13}},\ \bibinfo {pages} {013017}
  (\bibinfo {year} {2011})}\BibitemShut {NoStop}%
\bibitem [{EPA()}]{EPAPS}%
  \BibitemOpen
  \href@noop {} {}\bibinfo {note} {See the EPAPS for discussions on scattering
  matrix, finite temperature, finite bandwidth, internal loss, non-RWA
  correction, auxiliary cavity for phonon output, and the detailed
  characterization of tripartite entanglement.}\BibitemShut {Stop}%
\bibitem [{\citenamefont {Adesso}\ \emph {et~al.}(2012)\citenamefont {Adesso},
  \citenamefont {Girolami},\ and\ \citenamefont {Serafini}}]{Adesso2012}%
  \BibitemOpen
  \bibfield  {author} {\bibinfo {author} {\bibfnamefont {G.}~\bibnamefont
  {Adesso}}, \bibinfo {author} {\bibfnamefont {D.}~\bibnamefont {Girolami}}, \
  and\ \bibinfo {author} {\bibfnamefont {A.}~\bibnamefont {Serafini}},\ }\href
  {\doibase 10.1103/PhysRevLett.109.190502} {\bibfield  {journal} {\bibinfo
  {journal} {Phys. Rev. Lett.}\ }\textbf {\bibinfo {volume} {109}},\ \bibinfo
  {pages} {190502} (\bibinfo {year} {2012})}\BibitemShut {NoStop}%
\bibitem [{\citenamefont {Braunstein}\ and\ \citenamefont {van
  Loock}(2005)}]{Braunstein2005}%
  \BibitemOpen
  \bibfield  {author} {\bibinfo {author} {\bibfnamefont {S.}~\bibnamefont
  {Braunstein}}\ and\ \bibinfo {author} {\bibfnamefont {P.}~\bibnamefont {van
  Loock}},\ }\href@noop {} {\bibfield  {journal} {\bibinfo  {journal} {Rev.
  Mod. Phys.}\ }\textbf {\bibinfo {volume} {77}},\ \bibinfo {pages} {513}
  (\bibinfo {year} {2005})}\BibitemShut {NoStop}%
\bibitem [{\citenamefont {Koike}\ \emph {et~al.}(2006)\citenamefont {Koike},
  \citenamefont {Takahashi}, \citenamefont {Yonezawa}, \citenamefont {Takei},
  \citenamefont {Braunstein}, \citenamefont {Aoki},\ and\ \citenamefont
  {Furusawa}}]{Furusawa2006}%
  \BibitemOpen
  \bibfield  {author} {\bibinfo {author} {\bibfnamefont {S.}~\bibnamefont
  {Koike}}, \bibinfo {author} {\bibfnamefont {H.}~\bibnamefont {Takahashi}},
  \bibinfo {author} {\bibfnamefont {H.}~\bibnamefont {Yonezawa}}, \bibinfo
  {author} {\bibfnamefont {N.}~\bibnamefont {Takei}}, \bibinfo {author}
  {\bibfnamefont {S.~L.}\ \bibnamefont {Braunstein}}, \bibinfo {author}
  {\bibfnamefont {T.}~\bibnamefont {Aoki}}, \ and\ \bibinfo {author}
  {\bibfnamefont {A.}~\bibnamefont {Furusawa}},\ }\href@noop {} {\bibfield
  {journal} {\bibinfo  {journal} {Phys. Rev. Lett.}\ }\textbf {\bibinfo
  {volume} {96}} (\bibinfo {year} {2006})}\BibitemShut {NoStop}%
\bibitem [{\citenamefont {Bergeal}\ \emph {et~al.}(2010)\citenamefont
  {Bergeal}, \citenamefont {Vijay}, \citenamefont {Manucharyan}, \citenamefont
  {Siddiqi}, \citenamefont {Schoelkopf},\ and\ \citenamefont
  {Devoret}}]{Bergeal2010}%
  \BibitemOpen
  \bibfield  {author} {\bibinfo {author} {\bibfnamefont {N.}~\bibnamefont
  {Bergeal}}, \bibinfo {author} {\bibfnamefont {R.}~\bibnamefont {Vijay}},
  \bibinfo {author} {\bibfnamefont {V.~E.}\ \bibnamefont {Manucharyan}},
  \bibinfo {author} {\bibfnamefont {I.}~\bibnamefont {Siddiqi}}, \bibinfo
  {author} {\bibfnamefont {R.~J.}\ \bibnamefont {Schoelkopf}}, \ and\ \bibinfo
  {author} {\bibfnamefont {M.}~\bibnamefont {Devoret}},\ }\href@noop {}
  {\bibfield  {journal} {\bibinfo  {journal} {Nature Phys.}\ }\textbf {\bibinfo
  {volume} {296}} (\bibinfo {year} {2010})}\BibitemShut {NoStop}%
\bibitem [{Bau()}]{Baust2014}%
  \BibitemOpen
  \href@noop {} {}\bibinfo {note} {Baust, A. et. al.,
  arXiv:1405.1969}\BibitemShut {NoStop}%
\bibitem [{\citenamefont {Gardiner}\ and\ \citenamefont
  {Zoller}(2004)}]{Gardinerbook}%
  \BibitemOpen
  \bibfield  {author} {\bibinfo {author} {\bibfnamefont {C.}~\bibnamefont
  {Gardiner}}\ and\ \bibinfo {author} {\bibfnamefont {P.}~\bibnamefont
  {Zoller}},\ }\href@noop {} {\emph {\bibinfo {title} {Quantum Noise}}},\
  \bibinfo {edition} {3rd}\ ed.\ (\bibinfo  {publisher} {Springer},\ \bibinfo
  {address} {New York},\ \bibinfo {year} {2004})\BibitemShut {NoStop}%
\bibitem [{\citenamefont {Wang}\ and\ \citenamefont {Clerk}(2013)}]{Wang2013}%
  \BibitemOpen
  \bibfield  {author} {\bibinfo {author} {\bibfnamefont {Y.-D.}\ \bibnamefont
  {Wang}}\ and\ \bibinfo {author} {\bibfnamefont {A.~A.}\ \bibnamefont
  {Clerk}},\ }\href {\doibase 10.1103/PhysRevLett.110.253601} {\bibfield
  {journal} {\bibinfo  {journal} {Phys. Rev. Lett.}\ }\textbf {\bibinfo
  {volume} {110}},\ \bibinfo {pages} {253601} (\bibinfo {year}
  {2013})}\BibitemShut {NoStop}%
\bibitem [{\citenamefont {DeJesus}\ and\ \citenamefont
  {Kaufman}(1987)}]{DeJesus1987}%
  \BibitemOpen
  \bibfield  {author} {\bibinfo {author} {\bibfnamefont {E.~X.}\ \bibnamefont
  {DeJesus}}\ and\ \bibinfo {author} {\bibfnamefont {C.}~\bibnamefont
  {Kaufman}},\ }\href@noop {} {\bibfield  {journal} {\bibinfo  {journal} {Phys.
  Rev. A}\ }\textbf {\bibinfo {volume} {5288}},\ \bibinfo {pages} {35}
  (\bibinfo {year} {1987})}\BibitemShut {NoStop}%
\bibitem [{\citenamefont {Vidal}\ and\ \citenamefont
  {Werner}(2002)}]{Vidal2002}%
  \BibitemOpen
  \bibfield  {author} {\bibinfo {author} {\bibfnamefont {G.}~\bibnamefont
  {Vidal}}\ and\ \bibinfo {author} {\bibfnamefont {R.~F.}\ \bibnamefont
  {Werner}},\ }\href {\doibase 10.1103/PhysRevA.65.032314} {\bibfield
  {journal} {\bibinfo  {journal} {Phys. Rev. A}\ }\textbf {\bibinfo {volume}
  {65}},\ \bibinfo {pages} {032314} (\bibinfo {year} {2002})}\BibitemShut
  {NoStop}%
\bibitem [{\citenamefont {Plenio}(2005)}]{Plenio2005}%
  \BibitemOpen
  \bibfield  {author} {\bibinfo {author} {\bibfnamefont {M.~B.}\ \bibnamefont
  {Plenio}},\ }\href@noop {} {\bibfield  {journal} {\bibinfo  {journal} {Phys.
  Rev. Lett.}\ }\textbf {\bibinfo {volume} {95}},\ \bibinfo {pages} {090503}
  (\bibinfo {year} {2005})}\BibitemShut {NoStop}%
\bibitem [{\citenamefont {Walls}\ and\ \citenamefont
  {Milburn}(1994)}]{WallsMilburn}%
  \BibitemOpen
  \bibfield  {author} {\bibinfo {author} {\bibfnamefont {D.~F.}\ \bibnamefont
  {Walls}}\ and\ \bibinfo {author} {\bibfnamefont {G.~J.}\ \bibnamefont
  {Milburn}},\ }\href@noop {} {\emph {\bibinfo {title} {Quantum optics}}}\
  (\bibinfo  {publisher} {Springer, Berlin},\ \bibinfo {year}
  {1994})\BibitemShut {NoStop}%
\bibitem [{\citenamefont {Metelmann}\ and\ \citenamefont
  {Clerk}(2014)}]{Metelmann2014}%
  \BibitemOpen
  \bibfield  {author} {\bibinfo {author} {\bibfnamefont {A.}~\bibnamefont
  {Metelmann}}\ and\ \bibinfo {author} {\bibfnamefont {A.~A.}\ \bibnamefont
  {Clerk}},\ }\href@noop {} {\bibfield  {journal} {\bibinfo  {journal} {Phys.
  Rev. Lett.}\ }\textbf {\bibinfo {volume} {112}},\ \bibinfo {pages} {133904}
  (\bibinfo {year} {2014})}\BibitemShut {NoStop}%
\bibitem [{\citenamefont {Eichenfield}\ \emph {et~al.}(2009)\citenamefont
  {Eichenfield}, \citenamefont {Chan}, \citenamefont {Camacho}, \citenamefont
  {Vahala},\ and\ \citenamefont {Painter}}]{Eichenfield2009b}%
  \BibitemOpen
  \bibfield  {author} {\bibinfo {author} {\bibfnamefont {M.}~\bibnamefont
  {Eichenfield}}, \bibinfo {author} {\bibfnamefont {J.}~\bibnamefont {Chan}},
  \bibinfo {author} {\bibfnamefont {R.~M.}\ \bibnamefont {Camacho}}, \bibinfo
  {author} {\bibfnamefont {K.~J.}\ \bibnamefont {Vahala}}, \ and\ \bibinfo
  {author} {\bibfnamefont {O.}~\bibnamefont {Painter}},\ }\href@noop {}
  {\bibfield  {journal} {\bibinfo  {journal} {Nature}\ }\textbf {\bibinfo
  {volume} {461}},\ \bibinfo {pages} {78} (\bibinfo {year} {2009})}\BibitemShut
  {NoStop}%
\bibitem [{\citenamefont {Ferraro}\ \emph {et~al.}(2004)\citenamefont
  {Ferraro}, \citenamefont {Paris}, \citenamefont {Bondani}, \citenamefont
  {Allevi}, \citenamefont {Puddu},\ and\ \citenamefont
  {Andreoni}}]{Ferraro2004}%
  \BibitemOpen
  \bibfield  {author} {\bibinfo {author} {\bibfnamefont {A.}~\bibnamefont
  {Ferraro}}, \bibinfo {author} {\bibfnamefont {M.~G.~A.}\ \bibnamefont
  {Paris}}, \bibinfo {author} {\bibfnamefont {M.}~\bibnamefont {Bondani}},
  \bibinfo {author} {\bibfnamefont {A.}~\bibnamefont {Allevi}}, \bibinfo
  {author} {\bibfnamefont {E.}~\bibnamefont {Puddu}}, \ and\ \bibinfo {author}
  {\bibfnamefont {A.}~\bibnamefont {Andreoni}},\ }\href@noop {} {\bibfield
  {journal} {\bibinfo  {journal} {J. Opt. Soc. Am. B}\ }\textbf {\bibinfo
  {volume} {21}},\ \bibinfo {pages} {1241} (\bibinfo {year}
  {2004})}\BibitemShut {NoStop}%
\bibitem [{\citenamefont {Giedke}\ \emph {et~al.}(2001)\citenamefont {Giedke},
  \citenamefont {Kraus}, \citenamefont {Lewenstein},\ and\ \citenamefont
  {Cirac}}]{Giedke2001}%
  \BibitemOpen
  \bibfield  {author} {\bibinfo {author} {\bibfnamefont {G.}~\bibnamefont
  {Giedke}}, \bibinfo {author} {\bibfnamefont {B.}~\bibnamefont {Kraus}},
  \bibinfo {author} {\bibfnamefont {M.}~\bibnamefont {Lewenstein}}, \ and\
  \bibinfo {author} {\bibfnamefont {J.~I.}\ \bibnamefont {Cirac}},\ }\href@noop
  {} {\bibfield  {journal} {\bibinfo  {journal} {Phys. Rev. A}\ }\textbf
  {\bibinfo {volume} {64}},\ \bibinfo {pages} {052303} (\bibinfo {year}
  {2001})}\BibitemShut {NoStop}%
\bibitem [{\citenamefont {van Loock}\ and\ \citenamefont
  {Braunstein}(2000)}]{vanLoock2000}%
  \BibitemOpen
  \bibfield  {author} {\bibinfo {author} {\bibfnamefont {P.}~\bibnamefont {van
  Loock}}\ and\ \bibinfo {author} {\bibfnamefont {S.~L.}\ \bibnamefont
  {Braunstein}},\ }\href@noop {} {\bibfield  {journal} {\bibinfo  {journal}
  {Phys. Rev. Lett.}\ }\textbf {\bibinfo {volume} {84}},\ \bibinfo {pages}
  {3482} (\bibinfo {year} {2000})}\BibitemShut {NoStop}%
\end{thebibliography}

\begin{thebibliography}{3}
\expandafter\ifx\csname natexlab\endcsname\relax\def\natexlab#1{#1}\fi
\expandafter\ifx\csname bibnamefont\endcsname\relax
  \def\bibnamefont#1{#1}\fi
\expandafter\ifx\csname bibfnamefont\endcsname\relax
  \def\bibfnamefont#1{#1}\fi
\expandafter\ifx\csname citenamefont\endcsname\relax
  \def\citenamefont#1{#1}\fi
\expandafter\ifx\csname url\endcsname\relax
  \def\url#1{\texttt{#1}}\fi
\expandafter\ifx\csname urlprefix\endcsname\relax\def\urlprefix{URL }\fi
\providecommand{\bibinfo}[2]{#2}
\providecommand{\eprint}[2][]{\url{#2}}

\bibitem[{\citenamefont{Tian}(2013)}]{Tian2013supp}
\bibinfo{author}{\bibfnamefont{L.}~\bibnamefont{Tian}}, \bibinfo{journal}{Phys.
  Rev. Lett.} \textbf{\bibinfo{volume}{110}}, \bibinfo{pages}{233602}
  (\bibinfo{year}{2013}).

\bibitem[{\citenamefont{Wang and Clerk}(2013)}]{Wang2013supp}
\bibinfo{author}{\bibfnamefont{Y.-D.} \bibnamefont{Wang}} \bibnamefont{and}
  \bibinfo{author}{\bibfnamefont{A.~A.} \bibnamefont{Clerk}},
  \bibinfo{journal}{Phys. Rev. Lett.} \textbf{\bibinfo{volume}{110}},
  \bibinfo{pages}{253601} (\bibinfo{year}{2013}).

\bibitem[{\citenamefont{Adesso et~al.}(2012)\citenamefont{Adesso, Girolami, and
  Serafini}}]{Adesso2012supp}
\bibinfo{author}{\bibfnamefont{G.}~\bibnamefont{Adesso}},
  \bibinfo{author}{\bibfnamefont{D.}~\bibnamefont{Girolami}}, \bibnamefont{and}
  \bibinfo{author}{\bibfnamefont{A.}~\bibnamefont{Serafini}},
  \bibinfo{journal}{Phys. Rev. Lett.} \textbf{\bibinfo{volume}{109}},
  \bibinfo{pages}{190502} (\bibinfo{year}{2012}).

\end{thebibliography}
\end{document}